\documentclass[12pt]{iopart}
\usepackage{iopams}                            
\usepackage{subfigure}
\usepackage{graphicx}
\usepackage{color}
\usepackage[12hr]{datetime}                    
\usdate
\usepackage{subeqn}                            

\newcommand{\Schrodinger}{Schr{\"o}dinger}     
\newcommand{\PT}{\mathcal{PT}}                 
\newcommand{\mc}[1]{\mathcal{#1}}              

\newcommand{\ef}[1]{(\ref{#1})}                
\newcommand{\eg}{\textit{e.g.}}                
\newcommand{\ie}{\textit{i.e.}}                
\newcommand{\vs}{\textit{vs}}                  
\newcommand{\dd}[1]{\,\mathrm{d}#1\,}
\newcommand{\sech}{\,\mathrm{sech}}

\newcommand{\tfrac}[2]{\textstyle{\frac{#1}{#2}}}   
\newcommand{\tint}{\!\int\!}                   
\newcommand{\PoissonB}[2]{\ensuremath{ \lbrace \,#1,#2\, \rbrace }}  
\newcommand{\tpsi}{\tilde{\psi}}               
\newcommand{\tchi}{\tilde{\chi}}               
\newcommand{\dv}[2]{\frac{\mathrm{d}#1}{\mathrm{d} #2}}
\newcommand{\pdv}[2]{\frac{\partial #1}{\partial #2}}
\newcommand{\qc}{\>,\qquad}

\newcommand{\notag}{\nonumber}                 
\newcommand{\GammaF}[1]{\Gamma[\,#1\,]}

\newcommand{\be}{\begin{equation}}
\newcommand{\ee}{\end{equation}}
\newcommand{\bea}{\begin{eqnarray}}
\newcommand{\eea}{\end{eqnarray}}
\newcommand{\calF}{\cal{F}}
\newcommand{\fdv}[2]{\frac{\delta #1}{\delta #2}}
\eqnobysec                                     
\begin{document}
\title[cDNLSE equation]{Exact Solutions of a generalized variant of the derivative nonlinear \Schrodinger\ equation in a  Scarff II external potential and their stability properties}
\author{Avinash Khare$^1$, Fred Cooper$^{2,3}$, John F. Dawson$^4$ 
}
\address{$^1$Physics Department, 
   Savitribai Phule Pune University, 
   Pune 411007, India}
\address{$^2$The Santa Fe Institute, 
   1399 Hyde Park Road, 
   Santa Fe, NM 87501, United States of America}
\address{$^3$Theoretical Division, 
   Los Alamos National Laboratory, 
   Los Alamos, NM 87545, United States of America}
\address{$^4$Department of Physics, 
   University of New Hampshire, 
   Durham, NH 03824, United States of America}   
\vspace{10pt}
\begin{indented}
\item[]\today, \currenttime\ PST
\end{indented}
\begin{abstract}
We obtain exact solitary wave solutions of a variant of the generalized derivative nonlinear \Schrodinger\ equation in 1+1 dimensions with arbitrary values of the nonlinearity parameter $\kappa$ in a Scarf-II potential.  This variant of the usual derivative nonlinear \Schrodinger\ equation has the properties that for real external potentials, the dynamics is derivable from a Lagrangian.  The solitary wave and trapped  solutions have the same form as those of the usual derivative nonlinear \Schrodinger\ equation.  We show that the solitary wave solutions are orbitally stable for $\kappa \leq 1$  We find new exact nodeless solutions to the bound states in the external complex potential which are related to the static solutions of the  equation.  We also use a  collective coordinate approximation to analyze the stability of the trapped solutions when the external potential is real. 
\end{abstract}
\submitto{\jpa}
\vspace{2pc}
\noindent{\it Keywords}: exact solutions, stability analysis, collective coordinates, variational approach, dissipation functional. 
\maketitle
%
%
\section{\label{s:Intro}Introduction}

The derivative nonlinear \Schrodinger\ equation (DNLS equation) is an integrable system related to the nonlinear  \Schrodinger\ equation (NLSE).  It was first discussed by Kaup and Newell \cite{doi:10.1063/1.523737}. 
The partial differential equation for this system is 
\begin{equation}\label{IT.e:1}
\fl
   \rmi \psi_t + \psi_{xx} + \rmi \, g(|\psi|^{2} \psi)_x = 0 \>.
\end{equation}
Since its discovery by Kaup and Newell, the derivative NLSE has been studied by many people.  Among its many properties are the existence of stable moving solitary waves. Recently because of the interest of the behavior of solitary wave systems in the presence of complex external parity-time ($\PT$) symmetric potentials, there has been renewed interest in studying this equation in those environments which can be reproduced in optical fibers. 

The DNLS equation with general $\kappa$ power nonlinearity in the presence of an external $\PT$ symmetric Scarff-II potential \cite{Ahmed2001343} is given by
\begin{equation}\label{IT.e:2}
\fl
   \rmi \, \psi_t 
   + 
   \psi_{xx} 
   + 
   \rmi \, g( \, |\psi|^{2\kappa} \psi \, )_x 
   +
   [\, V_0 \sech^2 x + \rmi \, W_0 \sech x \tanh x \,] \, \psi(x)  
   =
   0 \>,
\end{equation}
with $V_0 > 0$.  This equation has recently been studied for the special case $\kappa=1$ by Chen and Yan \cite{PhysRevE.95.012205}. 
When $\kappa=1$,  this equation has been used to describe nonlinear waves in plasmas as well as optical fibers.  In Chen and Yan the regimes where the trapped solutions were stable was mapped out as a function of $(V_0,W_0)$.    

For the DNLS equation both static and moving solutions have been found at $\kappa=1$ and it has been proven that these solutions are stable as long as the velocity is less than a critical value.   The orbital  stability of these solutions when $\kappa=1$ and $V=0$ was shown by Colin and Ohta \cite{r:Colin:2006sh}. This was done by showing that the solutions of the DNLS equation could be mapped by a ``gauge transformation'' into solutions of a variant of the DNLS equation  which we denote the ``conserving" DNLS equation (cDNLSE)  which has a Lagrangian formulation: \ie\ the equation
\begin{equation}\label{IT.e:3}
\fl
   \rmi \, \psi_t 
   + 
   \psi_{xx} 
   + 
   \rmi \, g  \, |\psi|^{2} \psi_x \
     =
   0 \>,
\end{equation}
 can be derived from Lagrangian
\begin{eqnarray}
\fl
   L
   = 
   \tint \dd{x}
   \Bigl \{\,   
      \frac{\rmi}{2} 
      \bigl [\,
         \psi^{\ast}(x,t) \, \psi_t(x,t) 
         - 
         \psi^{\ast}_t(x,t) \, \psi(x,t) \,
      \bigr ]
      \label{IT.e:4} \\
      -
      | \psi_x(x,t)|^2 
      - 
      \frac{\rmi \, g}{4} 
        | \psi(x,t) |^{2} \, 
      [\, \psi^{\ast}(x,t) \, \psi_x(x,t) - \psi^{\ast}_x(x,t) \, \psi(x,t) 
   \Bigr \} \>, 
   \notag
\end{eqnarray}
so that the energy
\begin{equation}\label{IT.e:5}
\fl
   E
   =
   \tint \dd{x}
   \Bigl \{\,   
      | \psi_x(x,t)|^2 
      + 
      \frac{\rmi \, g}{4} 
        | \psi(x,t) |^{2} \, 
      [\, \psi^{\ast}(x,t) \, \psi_x(x,t) - \psi^{\ast}_x(x,t) \, \psi(x,t) 
   \Bigr \} 
\end{equation}
is conserved.

The transformation of the solutions of the DNLS equation, denoted by $v(x,t)$, to solutions of the cDNLS equation
which we denote by $\psi(x,t)$ is given by 
\be\label{IT.e:6}
\fl
   \psi(x,t) 
   = 
   v(x,t) \exp \left[ \frac{i}{2} \int_{-\infty}^x \!\! |v(x,t)|^2 \right] \>.
\ee
This gauge equivalency was used by Jenkins \etal \cite{Jenkins2018} to prove the global well-posedness of the solutions.  This transformation was also used to study the asymptotic (in time) of the traveling wave solutions by Hayashi and Ozawa \cite{HayashiOzawa1992}.
 
We were originally interested in studying the stability properties of the exact solutions  of the DNLS equation with and without a potential, when $\kappa \neq 1$ in analogy to what has been done for the nonlinear \Schrodinger\ equation with arbitrary $\kappa$.  However at arbitrary $\kappa$ we were unable to find a gauge transformation relating the original DNLS equation to the cDNLS equation.  
So instead we thought it preferable to first study the stability of the solutions
for arbitrary $\kappa$ power nonlinearity of the gauge transformed DNLS equation.  
This variant of the derivative NLSE is derivable from a Lagrangian when the external potential is real so that simple methods can be used to qualitatively describe stability properties.  A similar situation occurred in the study of compactons, where Cooper \etal \cite{COOPER1993344} modified the original equation of Camassa and Holm \cite{PhysRevLett.71.1661} so that it was derivable from a Lagrangian. 
The exact solutions of the generalizations the DNLS equation and the cDNLS equation to arbitrary $\kappa$ are quite similar in form which is analogous to what happened in the Camassa-Holm equation when compared to the Cooper-Shepard-Lucheeroni-Sodano (CSLS) equation.  So here, we will find exact solutions to the equation:
\begin{equation}\label{IT.e:7}
\fl
   \rmi \, \psi_t 
   + 
   \psi_{xx} 
   + 
   \rmi \, g \, |\psi|^{2\kappa} \psi_x 
   +
   [\, V_0 \sech^2 x + \rmi \, W_0 \sech x \tanh x \,] \, \psi(x)  
   =
   0 \>,
\end{equation}
for arbitrary $\kappa$, and study their stability as a function of $\kappa$ using a variety of methods when $W_0=0$. 
The exact solutions previously found for $\kappa=1$ can be easily  generalized to arbitrary $\kappa$, for both the DNLS equation and its variant, even when we include the complex external potential of the Scarff-II variety.

It is well known that for the nonlinear \Schrodinger\ equation with arbitrary $\kappa$, the stability properties depend on the power of the nonlinearity.  One has a transition from stable solutions to those that blow up at a critical value of the mass  $M = \int \dd{x} |\psi|^2$ at $\kappa=2$ and then all solutions are unstable for $\kappa >2 $. (See \eg\ Rose and Weinstein \cite{ROSE1988207}, Cooper, Shepard, and Lucheroni,  \cite{COOPER1992184}, Cooper, Shepard, Lucheroni, and Sodano \cite{COOPER1993344}).   When one adds an external potential that is confining, one can show for the NLSE that the stability regions increases as we increase the magnitude of the potential.  We utilized this approach in \cite{1751-8121-50-48-485205}
where all simple methods:  Derrick's theorem \cite{doi:10.1063/1.1704233} and variants of the Vakhitov-Kolokolov (V-K) stability criteria \cite{Vakhitov:1973aa} and the collective coordinate method yielded consistent results whose answers agreed with those found in numerical simulations. Here we will find that because of the extra derivative, the critical value of $\kappa$ will be $1$ when there is no potential present.
We obtain this by looking at Derrick's theorem  \cite{doi:10.1063/1.1704233} as well as the orbital stability at arbitrary $\kappa$. 
When we add a real attractive potential we will find that the nodeless bound solutions will have an increased domain of stability which depends on the strength of the potential. This will be done using Derrick's theorem as well as looking at the small oscillation equations in a collective coordinate approximation. 

This paper is organized as follows.  We will first generalize to arbitrary $\kappa$ the exact static and moving solutions that were were found previously in Ref.~\cite{PhysRevE.95.012205}. We show that the variant  DNLS equation  has similar solutions.  We then show that even in the presence of the Scarff-II potential, the equations have similar solutions.  We then utilize the conservation laws of the variant DNLS equation to discus the stability of the latter solutions with and without a real potential.  Finally we introduce a collective coordinate approach to study the dynamics of the perturbed solutions. 

%
%
\section{\label{s:Sol-gDNLS}Solutions of the  derivative nonlinear \Schrodinger\  equation and the conserved derivative nonlinear \Schrodinger\  equation at arbitrary nonlinearity parameter $\kappa$.}

First let us determine the solitary wave solutions of the derivative nonlinear \Schrodinger\ equation with arbitrary  $\kappa$
\begin{equation}\label{G.e:1}
   \rmi \, \partial_t \psi(x,t) 
   + 
   \partial_x^2 \psi(x,t) 
   + 
   \rmi g \, 
   \partial_x \, [\, |\psi(x,t)|^{2\kappa} \, \psi(x,t) \, ]
   =
   0 \>,  
\end{equation}
with arbitrary nonlinearity parameter $\kappa$.
In order to find solitary wave solutions to \ef{G.e:1}, we assume that the solution has the form
\begin{subeqnarray}\label{G.e:3}
   \psi(x,t) 
   &=
   u(x- c t) \, \rme^{ \rmi \, \chi(x,t) } \>,
   \label{G.e:3-a} \\
   \chi(x,t)
   &=
   \omega t 
   + 
   \frac{c}{2} \, (x - c t) 
   - 
   \alpha' g \int_{-\infty}^{x-ct} \!\! \dd{z} [u(z)]^{2\kappa} \>,
   \label{G.e:3-b}
\end{subeqnarray}
where $|\psi(x,t)|=u(x-c t)$, is only a function of $z = x - c t$ and $u(x-c t)$ is {\em real}.  Upon substitution of \ef{G.e:3} into \ef{G.e:1} we find
\begin{equation}\label{G.e:14}
   \alpha'
   =
   \alpha_1
   =
   \frac{2 \kappa +1}{2(\kappa+1)} \>= \frac{2+\gamma}{2(\gamma+1)} \>,
   \qquad
   \gamma = \frac{1}{\kappa} \>,
\end{equation}
and we obtain an equation for $u(z)$:
\begin{equation}\label{G.e:13}
   u''(z)
   +
   \Bigl [\,
      \frac{c^2}{4}
      -
      \omega 
      -
      \frac{c}{2} \, g \,  [u(z)]^{2\kappa} \,
      +
      \frac{2\kappa+1}{[2(\kappa+1)]^2} \, g^2 \, [u(z)]^{4\kappa} \,
   \Bigr ] \, u(z)
   =
   0 \>,
\end{equation}
When $\kappa=1$, this reduces to equation~(1.4) in Colin and Ohta \cite{r:Colin:2006sh}.
In order to find a solution of \ef{G.e:13}, we start from the ansatz:
\begin{equation}\label{G.e:15}
   u(z)
   =
   A \, [\, \cosh(\beta z) + B \,]^{-\frac{1}{2\kappa}} \>.
\end{equation}
After some algebra, we find that  $u(z)$ is given by
\begin{equation}\label{G.e:16}
\fl
   u(z)
   =
   \Bigl [\,
      \frac{(\kappa + 1) (4 \, \omega - c^2)}{ g \sqrt{4\,\omega}} \, 
   \Bigr ]^{\frac{1}{2 \kappa}} \,
   \Bigl [\, 
      \cosh(\kappa z \sqrt{4 \omega - c^2}) - \frac{c}{\sqrt{4 \, \omega}} \,
   \Bigr ]^{-\frac{1}{2 \kappa}} \>,
\end{equation}
and that 
\begin{equation}\label{GG.e:17}
\fl
   \chi(z, t)
   = 
   \omega t 
   + 
   \frac{c}{2} \, z 
   -
   \frac{2 \kappa +1}{\kappa} \,
   \tan ^{-1}
   \Bigl [\,
      \frac{c+2 \sqrt{\omega}}{\sqrt{4 \omega - c^2}}
      \tanh 
      \Bigl (\,
         \frac{1}{2} \kappa  z \sqrt{4 \omega - c^2} \,
      \Bigr ) \,
   \Bigr ] \>.
\end{equation}
Again we notice that at $\kappa = 1$, this reduces to equation~(1.3) in Colin and Ohta \cite{r:Colin:2006sh}.
Furthermore if we set $c=0$, we get the result that
\begin{equation}
\psi(x,t) = u(x) \, \rme^{ \rmi \, \chi(x,t) } \>,
\end{equation}
where 
\begin{equation}
   u(x) 
   =  
   \Bigl [\,
      (\kappa +1) \sqrt{2 \omega} \,
   \Bigr ]^{\frac{1}{2 \kappa}}
   \sech^{\frac{1}{2 \kappa }}
   \bigl (\,
      2 \kappa \sqrt{\omega } \, x \,
   \bigr ) \>,
\end{equation}
and
\begin{equation}
   \chi(x,t) 
   = 
   \omega t 
   -
   \frac{2 \kappa +1}{\kappa} \,
   \tan^{-1}
   \bigl ( 
      \tanh \bigl( \kappa \sqrt{\omega } z \bigr ) \,
   \bigr ) \>.
\end{equation}  
Static solutions exist for arbitrary values of the width parameter $\beta = 2 \kappa \sqrt{\omega}$, but only the $\beta=1$ solution smoothly goes into the nodeless trapped solutions when a trapping potential is present. 

It is amusing to note that form of the static solution for DNLS equation is of the  {\it same} form as the trapped solution found for the NLS plus the $\PT$-invariant Scarff II potential in \cite{PhysRevE.92.042901}.  One major difference between the solutions of the cDNLSE and those of  the NLSE is that in the latter case, the static solutions transform into the moving solution with a simple transformation $x \to x-c t$.  This is not the case for cDNLSE, so that the use of collective coordinates based on trial wave functions of the form $\sech^{\gamma/2}  (x-q(t)) $ that we use here for the trapped solution stability problem will not be appropriate for the problem of seeing what happens when a solitary wave travels through a confining potential.  For the latter problem we would use generalizations of the exact moving solutions which also can be treated analytically but requires Elliptic functions as seen in our discussion of orbital stability.

%
%
\subsection{\label{ss:cDNLSE}The conserved derivative nonlinear \Schrodinger\ equation} 

For the conserved derivative nonlinear \Schrodinger\ equation (cDNLSE) the equations of motion are
\begin{equation}\label{V.e:1}
   \phantom{-} \rmi \, \psi_t(x,t)
   +
   \psi_{xx}(x,t)
   +
   \rmi g \, | \psi(x,t) |^{2\kappa} \, \psi_x(x,t)
   =
   0 \>.
\end{equation}
If we assume the same ansatz  found in \ef{G.e:3}, 
\begin{subeqnarray}\label{V.e:2}
  \psi(x,t) 
   &=
   u(x- c t) \, \rme^{ \rmi \, \chi(x,t) } \>,
   \label{V.e:2-a} \\
   \chi(x,t)
   &=
   \omega t 
   + 
   \frac{c}{2} \, (x - c t) 
   - 
   \alpha' g \int_{-\infty}^{x-ct} \!\! \dd{z} [u(z)]^{2\kappa} \>,
   \label{V.e:2-b}
\end{subeqnarray}
where $u(x)$ is assumed to be {\it real}, we find that $u(z)$ obeys \ef{G.e:13} and so has the same form.  However we now find 
\begin{equation}\label{V.e:3}
   \alpha'
   =
   \alpha_2 
   = 
   \frac{1}{2(\kappa+1)}
   =
   \frac {\gamma} {2(1+\gamma)}
\end{equation}
so that 
\begin{equation}\label{V.e:4}
   \chi(z, t)
   = 
   \omega t 
   + 
   \frac{c}{2} \, z 
   -
   \frac{1}{\kappa} \, 
   \tan ^{-1}
   \Bigl [\,
      \frac{c+2 \sqrt{\omega}}{\sqrt{4 \omega - c^2}}
      \tanh 
      \Bigl (\,
         \frac{1}{2} \kappa  z \sqrt{4 \omega - c^2} \,
      \Bigr ) \,
   \Bigr ] \>.
\end{equation}
When $c=0$ we get the result
\begin{equation}\label{V.e:5}
\psi(x,t)   =   u(x) \, \rme^{ \rmi \, \chi(x,t) } \>,
\end{equation}
where now
\begin{equation}
   u(x) 
   =  
   \Bigl [\,
      (\kappa +1) \sqrt{2 \omega} \,
   \Bigr ]^{\frac{1}{2 \kappa}}
   \sech^{\frac{1}{2 \kappa }}
   \bigl (\,
      2 \kappa \sqrt{\omega } \, x \,
   \bigr ) \>,
\end{equation}
and
\begin{equation}
   \chi(x,t) 
   = 
   \omega t 
   -
   \frac{1}{\kappa} \,
   \tan^{-1}
   \bigl ( 
      \tanh \bigl( \kappa \sqrt{\omega } z \bigr ) \,
   \bigr ) \>.
\end{equation}  

%
%
\subsection{\label{ss:ConsLaws-gDNLS}Conservation Laws of the DNLS and cDNLS equations}

For $\kappa=1$ it is well known (see for example ref.~\cite{r:Colin:2006sh}) that the DNLS equation can be transformed by a gauge transformation and that it obeys the conservation laws of mass, momentum and energy, namely 
\begin{subeqnarray}\label{G.e:17}
\fl
   M
   &=
   \tint \dd{x} | \psi(x,t) |^2 \>,
   \label{G.e:17-a} \\
\fl
   P
   &= 
   \tint \dd{x}
   \Bigl \{\,
      \frac{i}{2}
      \bigl [\,
         \psi(x,t) \, \psi_x^{\ast}(x,t)
         - 
         \psi^{\ast}(x,t) \, \psi_x(x,t) \,
      \bigr ] 
      + 
      \frac{1}{2} \, 
      \bigl | \psi(x,t) \, \bigr |^4  \,
   \Bigr \} \>,  
   \label{G.e:17-b} \\
\fl
   E 
   &= 
   \tint \dd{x} 
   \Bigl \{\,  
      | \psi_x(x,t) |^2 
      + 
      \frac{3}{4 i } \, \bigl | \psi(x,t) \, \bigr |^2 \,
      \bigl [\, 
         \psi^{\ast}(x,t) \, \psi_x (x,t)
         - 
         \psi(x,t)\, \psi^{\ast}_x(x,t) \,
      \bigr ]
      \label{G.e:17-c} \\
\fl
      & \hspace{4em}
      + 
      \frac{1}{2} \bigl |\, \psi(x,t) \, \bigr |^4 \,
   \Bigr \}
   \notag
\end{subeqnarray}
are all conserved.  
However for arbitrary $\kappa$ there is no gauge transformation available, and we instead find that only 
\begin{subeqnarray}\label{G.e:18}
\fl
   M
   &=
   \tint \dd{x} | \psi(x,t) |^2 \>,
   \label{G.e:18-a} \\
\fl
   P
   &= 
   \tint \dd{x}
   \Bigl \{\,
      \frac{ i}{2}
      \bigl [\,
         \psi(x,t) \, \psi_x^{\ast}(x,t)
         - 
         \psi^{\ast}(x,t) \, \psi_x(x,t) \,
      \bigr ] 
      + 
      \frac{1}{\kappa + 1} \, 
      \bigl | \psi(x,t) \, \bigr |^{2\kappa + 2}  \,
   \Bigr \}
   \label{G.e:18-b}
\end{subeqnarray}
are conserved.

For the cDNLS equation, we have for arbitrary $\kappa$ that energy is conserved. This is because the equation can be derived from an action principle.  The action for the cDNLS equation is
\begin{subeqnarray}\label{V.e:7}
\fl
   \Gamma[\psi,\psi^{\ast}]
   &=
   \tint \dd{t} \{\, T[\psi,\psi^{\ast}] - H[\psi,\psi^{\ast}] \, \} \>,
   \quad
   H[\psi,\psi^{\ast}]
   =
   H_1[\psi,\psi^{\ast}] - H_2[\psi,\psi^{\ast}]
   \label{V.e:7-a} \\
\fl
   T[\psi,\psi^{\ast}]
   &=
   \frac{\rmi}{2} \tint \dd{x}  
   [\, \psi^{\ast}(x,t) \, \psi_t(x,t) - \psi^{\ast}_t(x,t) \, \psi(x,t) \,] \>,
   \label{V.e:7-b} \\
\fl
   H_1[\psi,\psi^{\ast}]
   &=
   \tint \dd{x} | \psi_x(x,t) |^2 \>,
   \label{V.e:7-c} \\
\fl
   H_2[\psi,\psi^{\ast}]
   &=
   \frac{\rmi \, g}{2(\kappa+1)} 
   \tint \dd{x} \,
   | \psi(x,t) |^{2 \kappa} \,
   [\, \psi^{\ast}(x,t) \, \psi_x(x,t) - \psi^{\ast}_x(x,t) \, \psi(x,t) \, ] \,
   \label{V.e:7-d}
\end{subeqnarray}
The  equations of motion arise from the condition that the action is stationary
\begin{equation}\label{V.e:8}
   \fdv{\Gamma[\psi,\psi^{\ast}]}{\psi^{\ast}(x,t)} = 0 \>,
   \qquad
   \fdv{\Gamma[\psi,\psi^{\ast}]}{\psi(x,t)} = 0 \>,
\end{equation}
which
reproduce \ef{V.e:1} and it's complex conjugate.  The equations of motion can also be obtained from Hamilton's equations:
\begin{subeqnarray}\label{V.e:9}
   \phantom{-}\rmi \, \psi_t(x,t)
   &=
   \frac{\delta H[\psi,\psi^{\ast}]}{\delta \psi^{\ast}(x,t)}
   =
   - 
   \psi_{xx}(x,t)
   -
   \rmi \, g \, | \psi(x,t) |^{2 \kappa} \, \psi_x(x,t) \>,
   \label{V.e:9-a} \\
   -\rmi \, \psi_t^{\ast}(x,t)
   &=
   \frac{\delta H[\psi,\psi^{\ast}]}{\delta \psi(x,t)}
   =
   - 
   \psi_{xx}^{\ast}(x,t)
   +
   \rmi \, g \, | \psi(x,t) |^{2 \kappa} \, \psi_x^{\ast}(x,t) \>.
   \label{V.e:9-b}
\end{subeqnarray}
This means that the Hamiltonian is conserved.  In addition, the mass (normalization) and the momentum are conserved:
\begin{subeqnarray}\label{MP}
   M[\psi,\psi^{\ast}]
   &=
   \tint \dd{x} | \psi(x,t) |^2 \>,
   \label{V.e:10-a} \\
   P[\psi,\psi^{\ast}]
   &=
   \frac{1}{2 \rmi} 
   \tint \dd{x} [\, 
      \psi^{\ast}(x,t) \, \psi_x(x,t) 
      - 
      \psi^{\ast}_x(x,t) \, \psi(x,t) \, ] \>.
   \label{V.e:10-b}   
\end{subeqnarray}
Introducing the real parameters $\omega$ and $c$ by means of the substitution,
\begin{equation}\label{V.e:11}
   \psi(x,t)
   =
   \phi(x - c t) \, \rme^{\rmi \omega t} \>,
   \qquad
   \psi^{\ast}(x,t)
   =
   \phi^{\ast}(x - c t) \, \rme^{-\rmi \omega t} \>, 
\end{equation}
The equations of motion \ef{V.e:9} become:
\begin{subeqnarray}\label{Ham}
\fl
   {}-
   \omega \, \phi(z)^{\phantom\ast} 
   - \,
   \rmi \, c \, \phi'(z)
   &=
   \frac{\delta H[\phi,\phi^{\ast}]}{\delta \phi^{\ast}(z)}
   =
   - 
   \phi''(z)\,
   -
   \rmi \, g \, | \phi(z) |^{2 \kappa} \, \phi'(z) \>,
   \label{V.e:12-a} \\
\fl
   {}-
   \omega \, \phi^{\ast}(z)
   + 
   \rmi \, c \, \phi^{\prime \ast}(z)
   &=
   \frac{\delta H[\phi,\phi^{\ast}]}{\delta \phi(z)}
   =
   - 
   \phi^{\prime\prime \ast}(z)
   +
   \rmi \, g \, | \phi(z) |^{2 \kappa} \, \phi^{\prime \ast}(z) \>,   
   \label{V.e:12-b}   
\end{subeqnarray}
where we have set $z = x - \omega t$, and a prime designates a derivative with respect to the argument.  The exact solutions $\phi(z)$ to \ef{Ham} depend on $\omega$ and $c$.
The conserved quantities now become
\begin{subeqnarray}\label{V.e:13}
\fl
   M[\phi,\phi^{\ast}]
   &=
   \tint \dd{z} | \phi(z) |^2 \>,
   \label{V.e:13-a} \\
\fl
   P[\phi,\phi^{\ast}]
   &=
   \frac{1}{2 \rmi} 
   \tint \dd{z} [\, 
      \phi^{\ast}(z) \, \phi'(z) 
      - 
      \phi^{\ast\prime}(z) \, \phi(z) \, ] \>.
   \label{V.e:13-b} \\
\fl
   E[\phi,\phi^{\ast}]
   &=
   \tint \dd{z} \,
   \Bigl \{\,
      | \phi'(z) |^2
      -
      \frac{\rmi \, g}{2(\kappa+1)} \, | \phi(z) |^{2 \kappa} \,
      [\, \phi^{\ast}(z) \, \phi'(z) 
         - 
         \phi^{\prime \ast}(z) \, \phi(z) \, ] \,
   \Bigr \} \>.
   \label{V.e:13-c}
\end{subeqnarray}

%
%
\subsection{\label{ss:VarPrinc}Variational Principle for Orbital stability}

We show  now that the equations for $\phi(z)$ and $\phi^\star(z)$ given in \ef{Ham} can be obtained via a variational principle by requiring the energy functional to be a minimum subject to the  constraints that the mass $M[\phi,\phi^{\ast}] = M_0$ and momentum $P[\phi,\phi^{\ast}] = P_0$ are fixed.  Introducing Lagrange multipliers $\omega$ and $c$, this is equivalent to minimizing the functional,
\begin{equation}\label{V.e:14}
   S_{\omega,c}[\phi,\phi^{\ast}]
   =
   E[\phi,\phi^{\ast}]
   +
   \omega \, M[\phi,\phi^{\ast}]
   -
   c \, P[\phi,\phi^{\ast}] \>.
\end{equation}
Noting that
\begin{eqnarray}\label{V.e:15}
   \phantom{-} \frac{\delta M[\phi,\phi^{\ast}]}{\delta \phi^{\ast}(z)}
   &=
   \phi(z) \>,
   \qquad
   \>\> \phantom{-} \frac{\delta M[\phi,\phi^{\ast}]}{\delta \phi(z)}
   =
   \phi^{\ast}(z) \>,
   \\
   -\frac{\delta P[\phi,\phi^{\ast}]}{\delta \phi^{\ast}(z)}
   &=
   \rmi \, \phi'(z) \>,
   \qquad
   -\frac{\delta P[\phi,\phi^{\ast}]}{\delta \phi(z)}
   =
   - \rmi \, \phi^{\prime\ast}(z) \>,
   \notag
\end{eqnarray}
we see that Hamilton's equations in the form given in \ef{Ham} are reproduced by requiring:
\begin{eqnarray}\label{V.e:16}
   \frac{\delta S_{\omega,c}[\phi,\phi^{\ast}]}
   {\delta \phi^{\ast}(z)} \, \bigg |_{\phi,\phi^{\ast}}
   &=
   0 \>,
   \qquad
   \frac{\delta S_{\omega,c}[\phi,\phi^{\ast}]}
   {\delta \phi(z)}\, \bigg |_{\phi,\phi^{\ast}}
   =
   0 \>.
\end{eqnarray}
We now follow Colin and Ohta \cite{r:Colin:2006sh} and define the function 
\begin{equation}
   d(\omega,c) = S_{\omega,c}( \phi_{\omega,c})
\end{equation}
where $\phi_{\omega,c}$ is the exact traveling solitary wave solution.  They have shown that if 
\begin{equation}
   \det[\, d''(\omega,c) \,] < 0 \>,
\end{equation}
then the solitary wave is orbitally stable.  To determine the stability we therefore need to calculate the determinant of the matrix
\begin{equation}\label{e:ddmatrix}
   dd[\omega,c,\kappa]
   =
   \Biggl (
   \begin{array}{cc}
      P_c[\omega,c,\kappa] & P_\omega[\omega,c,\kappa] \\
      M_c[\omega,c,\kappa] & M_\omega[\omega,c,\kappa]
   \end{array}
   \Biggr )
\end{equation}
where $P$ and $M$ are evaluated for the exact solution and the subscripts denote the partial derivatives with respect to $\omega$ or $c$.  Letting
\begin{equation}
    \alpha = (4 \omega-c^2) \, \kappa^2 > 0 \>,
    \qquad
    \sigma = c/(2 \sqrt \omega) \>,
\end{equation}
then 
\begin{equation}
   \phi_{\omega,c}^2(z) 
   =  
   \left[\,
      \frac{(\kappa+1) \alpha}{2 \kappa^2 \sqrt{\omega} }\,
   \right]^{\frac{1}{\kappa}}
   \left[\,
      \cosh \sqrt{\alpha} z-\sigma \,
   \right]^{-\frac{1}{\kappa}}
\end{equation}
For $M[\omega, c, \kappa]$ we need to calculate
\begin{equation}
\fl
   M[\omega, c, \kappa] 
   = 
   2 \int_0^\infty \!\! \dd{z}
   \phi_{\omega,c}^2(z) 
   =  
   \frac{2}{\sqrt \alpha} \,
      \left[\,
      \frac{(\kappa+1) \alpha}{2 \kappa^2 \sqrt{\omega} }\,
   \right]^{\frac{1}{\kappa}}
   \int_0^\infty \!\! \dd{y}
   \bigl [\, \cosh y -\sigma \, \bigr ]^{-\frac{1}{\kappa}} \>.
\end{equation}
So we need the integral 
\begin{eqnarray}
   I_1[\sigma, \kappa] 
   &=  
   \int_0^\infty \!\! \dd{y}
   \bigl [\, \cosh y -\sigma \,\bigr]^{-\frac{1}{\kappa}}
   \label{e:I1int} \\
   &=
   2^{\frac{1}{\kappa }} \kappa \,
   F_1
   \Bigl [\,
      \frac{1}{\kappa };
      \frac{1}{\kappa},
      \frac{1}{\kappa };
      1+\frac{1}{\kappa };
      \sigma +\rmi \, \sqrt{1-\sigma ^2},
      \sigma -\rmi \, \sqrt{1-\sigma ^2} \,
   \Bigr ] \>,
   \notag
\end{eqnarray}
which can be done analytically.  
Some particular cases are
\begin{subeqnarray}\label{e:I1v}
   I_1[\sigma,1]
   &=
   \frac{2}{\sqrt{1 - \sigma^2}} \,
   \tan^{-1}\Biggl [\, \sqrt{ \frac{1 + \sigma}{1 - \sigma} } \,\Biggr] \>,
   \label{e:I1v-a} \\
   I_1[\sigma,\tfrac{1}{2}]
   &=
   \frac{1}{(1 - \sigma^2)^{3/2}} \,
   \Bigl \{\,
      \sqrt{1 - \sigma^2}
      +
      2 \, \sigma \, 
      \tan^{-1}\Biggl [\, \sqrt{ \frac{1 + \sigma}{1 - \sigma} } \,\Biggr] \, 
   \Bigr \} \>,
   \label{e:I1v-b} \\
   I_1[\sigma,2]
   &=
   \frac{2}{\sqrt{1-\sigma}} \, 
   K \Bigl [\,\frac{\sigma + 1}{\sigma - 1} \, \Bigr ] \>,
   \label{e:I1v-c}
\end{subeqnarray}
where $K[x]$ is the elliptic integral of the first kind.  
This gives:
\begin{subeqnarray}\label{e:Mv}
   M[\omega,\sigma,1]
   &=
   8 \tan^{-1}\Biggl[\,\sqrt{\frac{1 + \sigma}{1 - \sigma}} \,\Biggr] \>,
   \label{e:Mv-a} \\
   M[\omega,\sigma,\tfrac{1}{2}]
   &=
   18 \sqrt{\omega }\,
   \Bigl \{\,
      \sqrt{1-\sigma ^2}
      +
      2 \, \sigma \,
      \tan^{-1}\Biggl[\,\sqrt{\frac{1 + \sigma}{1 - \sigma}} \,\Biggr] \,
   \Bigr \}
   \label{e:Mv-b} \\
   M[\omega,\sigma,2]
   &=
   \sqrt{ \frac{6}{\sqrt{\omega} \, (1 - \sigma)} } \,
   K \Bigl [\,\frac{\sigma + 1}{\sigma - 1} \, \Bigr ] \>.
   \label{e:Mv-c}   
\end{subeqnarray}
For the momentum $P$ we have that 
\begin{eqnarray}
\fl
   &P[\omega,c,\kappa]
   =
   -
   \int_{0}^{\infty} \!\!\!\! \dd{z} \phi_{\omega,c}^2(z) \,
   \Biggl [\,
      c - \frac{ \phi_{\omega,c}^{2\kappa}(z) }{ \kappa + 1} \,
   \Biggr ]
   \label{e:Pgeneq} \\
\fl
   &
   \quad
   =
   -
   \Biggl [\,
     \frac{(\kappa + 1)\,\alpha}{2 \kappa^2 \sqrt{\omega}} \,
   \Biggr ]^{\frac{1}{\kappa}} \,
   \int_{0}^{\infty} \!\!\!\! \dd{z}
   \bigl [\,
      \cosh(\sqrt{\alpha} z) - \sigma \,
   \bigr ]^{-\frac{1}{\kappa}} \,
   \Biggl [\,
      c
      -
      \frac{\alpha}{2 \kappa^2 \sqrt{\omega}} \,
      \bigl [\,
         \cosh(\sqrt{\alpha} z) - \sigma \,
      \bigr ]^{-1} \,
   \Biggr ] \>.
   \notag
\end{eqnarray}
After a change of variables to $y = z \sqrt \alpha$, we find the second integral in $P$ is related to 
\begin{equation}\label{e:I2def}
   I_2[\sigma, \kappa]
   =
   \int_{0}^{\infty} \!\!\!\! \dd{y}
   \bigl [\,
      \cosh y - \sigma \,
   \bigr ]^{-\bigl (\frac{1}{\kappa} + 1 \bigr)}
   =
   \kappa \dv{}{\sigma} \, I_1[\sigma,\kappa] \>,
\end{equation}
so that we can write $P$ in the form:
\begin{equation}\label{e:PI1I2}
   P[\omega,c,\kappa]
   =
   -
   \frac{1}{\sqrt{\alpha}} \,
   \Biggl [\,
     \frac{(\kappa + 1)\,\alpha}{2 \kappa^2 \sqrt{\omega}} \,
   \Biggr ]^{\frac{1}{\kappa}} \,
   \Biggl [\,
      c \, I_1[\sigma,\kappa]
      -
      \frac{\alpha}{2 \kappa^2 \sqrt{\omega}} \, I_2[\sigma,\kappa] \,
   \Biggr ] \>.
\end{equation}
Here $I_2[\sigma, \kappa]$ is given by
\begin{eqnarray}
\fl
   &I_2[\sigma, \kappa]
   =
   \frac{2^{\frac{1}{\kappa }}\,\kappa}{(\kappa +1) \sqrt{1-\sigma^2}} 
   \times
   \label{e:I2form} \\
\fl
   & \quad
   =
   \Biggl \{\,
      \Bigl (\,
         \sqrt{1 - \sigma^2} - \rmi \sigma \,
      \Bigr ) \,
         F_1
   \Bigl [\,
      1 + \frac{1}{\kappa };
      1 + \frac{1}{\kappa},
      \frac{1}{\kappa };
      2 + \frac{1}{\kappa };
      \sigma +\rmi \, \sqrt{1-\sigma ^2},
      \sigma -\rmi \, \sqrt{1-\sigma ^2} \,
   \Bigr ]
   \notag \\
   \fl
   & \qquad
   +
      \Bigl (\,
         \sqrt{1 - \sigma^2} + \rmi \sigma \,
      \Bigr ) \,
         F_1
   \Bigl [\,
      1 + \frac{1}{\kappa };
      \frac{1}{\kappa},
      1 + \frac{1}{\kappa };
      2 + \frac{1}{\kappa };
      \sigma +\rmi \, \sqrt{1-\sigma ^2},
      \sigma -\rmi \, \sqrt{1-\sigma ^2} \,
   \Bigr ]   
   \Biggr \} \>.
   \notag
\end{eqnarray}
Some particular cases are
\begin{subeqnarray}\label{e:I2v}
   I_2[\sigma,1]
   &=
   \frac{2}{(1 - \sigma^2)^{3/2}} \,
   \Bigl \{\,
      \sqrt{1 - \sigma^2}
      +
      2 \, \sigma 
      \tan^{-1}\Biggl[\,\sqrt{\frac{1 + \sigma}{1 - \sigma}} \,\Biggr] \,
   \Bigr \} \>,
   \label{e:I2v-a} \\
   I_2[\sigma,\tfrac{1}{2}]
   &=
   \frac{3 \, \sigma}{2 ( \sigma^2 - 1)^2}
   +
   \frac{1}{(1 - \sigma^2 )^{5/2}} \,
   \tan^{-1}\Biggl[\,\sqrt{\frac{1 + \sigma}{1 - \sigma}} \,\Biggr] \>,
   \label{e:I2v-b} \\
   I_2[\sigma,2]
   &=
   \frac{2}{(1 + \sigma) \sqrt{1-\sigma}} \,
   \Bigl \{ \,
      E \Bigl [\,\frac{\sigma + 1}{\sigma - 1} \, \Bigr ]
      -
      K \Bigl [\,\frac{\sigma + 1}{\sigma - 1} \, \Bigr ] \,
   \Bigr \} \>,
   \label{e:I2v-c}
\end{subeqnarray}
where $E[x]$ is the elliptic integral of the second kind.
This gives:
\begin{subeqnarray}\label{e:Pv}
   P[\omega,\sigma,1]
   &=
   4 \sqrt{\omega \, (1 - \sigma^2) } \>,
   \label{e:Pv-a} \\
   P[\omega,\sigma,\tfrac{1}{2}]
   &=
   9 \, \omega \,
   \Bigl \{\,
      \sigma \sqrt{1 - \sigma^2}
      +
      2 
      \tan^{-1}\Biggl[\,\sqrt{\frac{1 + \sigma}{1 - \sigma}} \,\Biggr] \,
   \Bigr \} \>,
   \label{e:Pv-b} \\
   P[\omega,\sigma,2]
   &=
   - 
   \sqrt{\frac{6 \sqrt{\omega}}{1 - \sigma}} \,
   \Bigl \{\,
      K \Bigl [\,\frac{\sigma + 1}{\sigma - 1} \, \Bigr ]
      +
      (\sigma - 1) \,
      E \Bigl [\,\frac{\sigma + 1}{\sigma - 1} \, \Bigr ] \,
   \Bigr \} \>.
   \label{e:Pv-c}
\end{subeqnarray}

%
%
\subsection{\label{ss:SpecialCases}Special Cases}

For $\kappa=1$, the matrix \ef{e:ddmatrix} of partial derivatives is given by
\begin{equation}\label{e:ddk1}
   dd[\omega,\sigma,1]
   =
   \frac{2}{\sqrt{1 - \omega^2}} \,
   \Biggl (
   \begin{array}{cc}
      - \sigma & 1/\sqrt{\omega} \\
      1/\sqrt{\omega} & - \sigma / \omega 
   \end{array}
   \Biggr )  \>,
\end{equation}
and the determinant is given by
\begin{equation}\label{e:detddk1}
   \det[\, dd[\omega,\sigma,1] \,]
   =
   - \frac{4}{\omega} < 0 \>,
\end{equation}
since $\omega > 0$.  Therefore we get the same result as Colin-Ohta \cite{r:Colin:2006sh} that the solitary wave is orbitally stable at $\kappa=1$. 

For $\kappa=1/2$, we expect the solitary waves to be stable. 
Here again the mass $M$ and momentum $P$ is given in terms of elementary functions, and we find
\begin{equation}\label{e:detddk1-2}
   \det[\, dd[\omega,\sigma,\tfrac{1}{2}] \,]
   =
   81 \,
   \bigl \{\,
      1
      - 
      \sigma^2
      +
      \bigl [\,
         \ln \bigl ( - \sigma + \rmi \sqrt{1 - \sigma^2} \, \bigr ) \,
      \bigr ]^2 \,
   \bigr \}
\end{equation}
which again is always negative for $0 < \sigma < 1$ showing stability. 

When $\kappa=2$, we expect the solution to be unstable and we find
\begin{equation}\label{e:detddk2}
\fl
   \det[\, dd[\omega,\sigma,2] \,]
   =
   - 
   \frac{\displaystyle{3 E \Bigl [\,\frac{\sigma + 1}{\sigma - 1} \, \Bigr ] }}
        {8 \, ( \sigma^2 - 1 ) \, \omega^{3/2}} \,
   \Bigl \{\,
      2 \, K \Bigl [\,\frac{\sigma + 1}{\sigma - 1} \, \Bigr ]
      +
      ( \sigma - 1 ) \,
      E \Bigl [\,\frac{\sigma + 1}{\sigma - 1} \, \Bigr ] \,
   \Bigr \} \>.
\end{equation}
So for $0 < \sigma < 1$, we find numerically that
\begin{equation}\label{e:detddk2limit}
   \det[\, dd[\omega,\sigma,2] \,]
   \ge
   \frac{1}{2 \, \omega^{3/2}}
   >
   0 \>,
\end{equation}
which shows that these solitary waves are orbitally unstable.

%
\subsection{\label{GenProps}General Properties of stability as a function of $\kappa$, $\sigma$}

\begin{figure}[t]
  \centering
  \subfigure[\ $\det\lbrack dd\lbrack\omega,\sigma,1.1\rbrack \rbrack$ \vs\ $\sigma$ when $\kappa=1.1$ \,]
  { \label{f:k11} 
  \includegraphics[width=0.45\columnwidth]{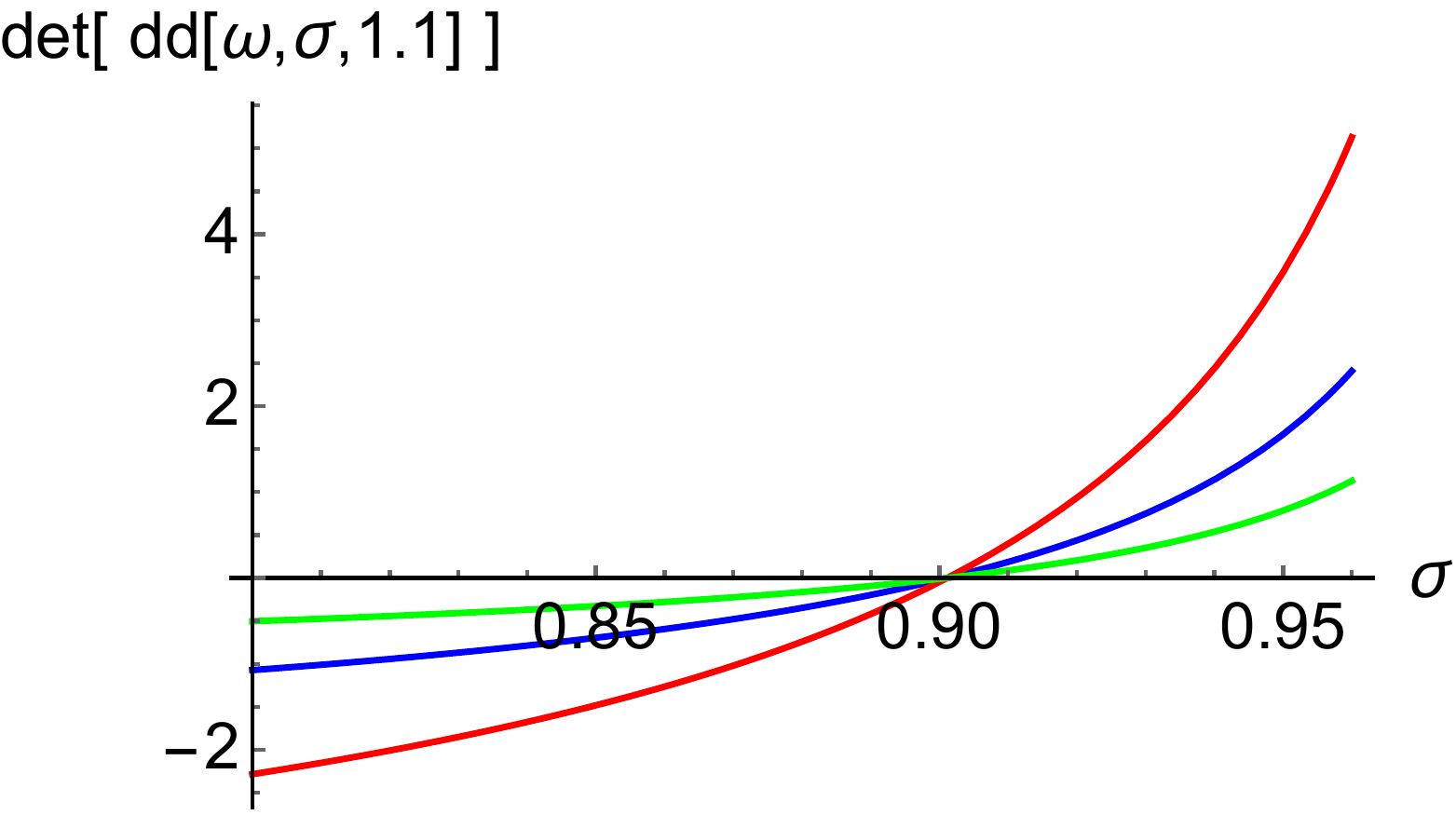} }
  \subfigure[\ $\det\lbrack dd\lbrack\omega,0,\kappa\rbrack \rbrack$ \vs\ $\kappa$ when $\sigma=0$ \,]
  { \label{f:sigzero}
  \includegraphics[width=0.45\columnwidth]{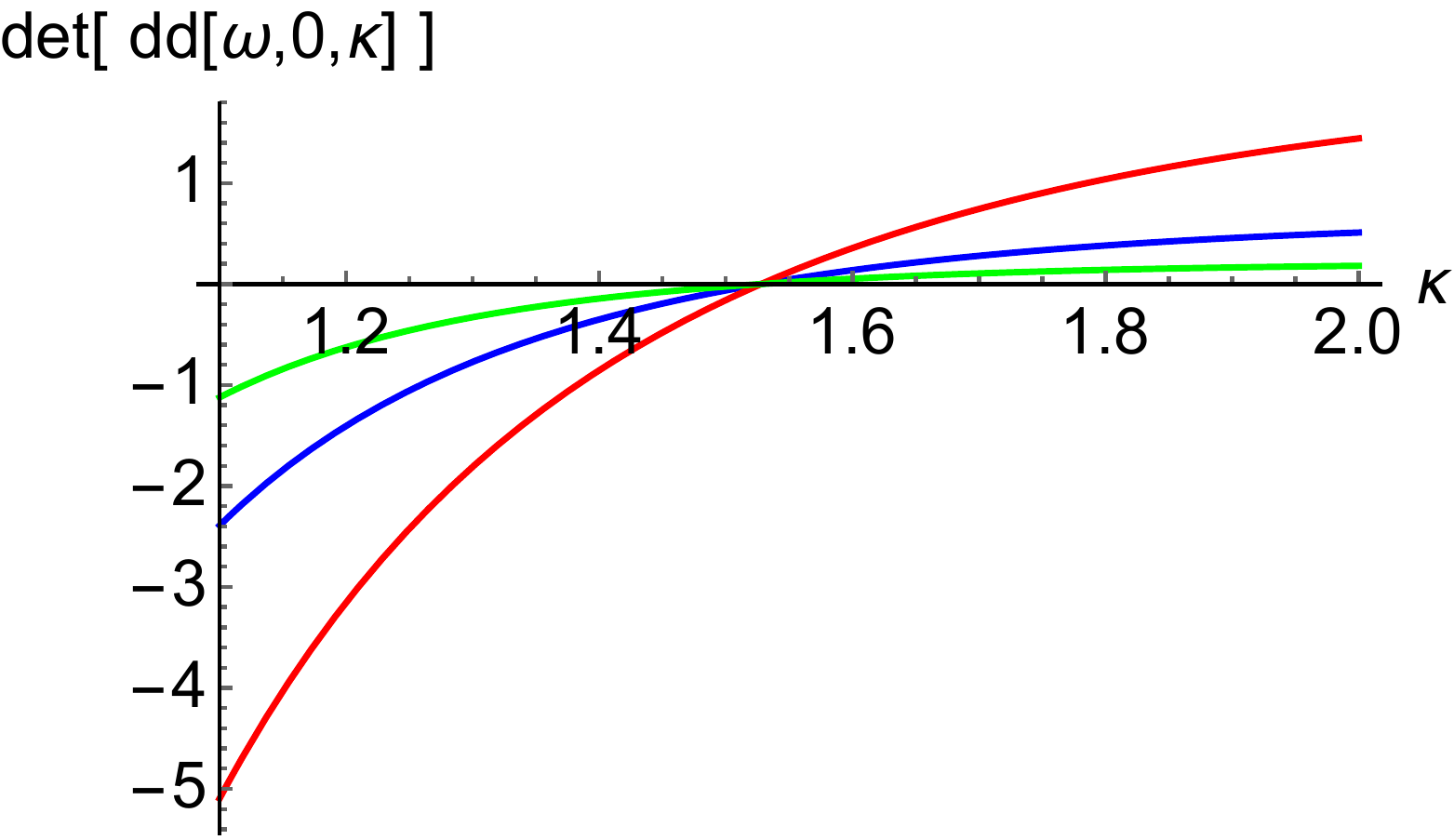} }
  \caption{(a) The determinant $\det[ dd[\omega,\sigma,\kappa] ]$ when $\kappa=1.1$ for the cases when
  $\omega = 1$ (blue), $\omega = 0.5$ (red), and $\omega = 2$ (green). (b) The determinant 
  $\det[ dd[\omega,\sigma,\kappa] ]$ as a function of $\kappa$ when $\sigma=0$.}
\end{figure}

As we will also find out from Derrick's theorem, $\kappa=1$ is the critical value of $\kappa$ below which all the soltuion for arbitrary allowed
$\sigma = c/(2 \sqrt \omega) < 1$ are stable.  Once $\kappa >1$, we find that the orbital stability first breaks down when 
 $\sigma = c/(2 \sqrt \omega)$ is close to one. For example,   at $\kappa=1.1$ the value of the determinant becomes positive at $\sigma \approx 0.9$. This is seen Fig.~\ref{f:k11} where we plot $\det[dd[1,\sigma,1.1]$ at $\omega=1$ as a function of $\sigma$.   After the completion of this work we discovered that  orbital stability of this model was previously studied by Liu, Simpson and Sulem \cite{Liu2013} .  Their more rigorous discussion, however,  relied on general properties of $I_1 [\sigma,\kappa]$.  Our analysis benefitted from having an explicit expression of $I_1$, which allowed us to determine the crossover to unstable behavior at arbitrary $\kappa$ as shown in Fig.~\ref{f:k11}.
%
%
%
%
The stable regime when $\kappa >1$ is a decreasing function of  $\sigma$.  There is a smaller and smaller  domain of stability as a function of $\sigma$ as one increases $\kappa$  until $\kappa \approx 1.528$, independent of $\omega$.  
This crossover is seen by plotting the determinant as a function of $\kappa$ at $\sigma=0$ as shown in Fig.~\ref{f:sigzero}
for different value of $\omega$.

%
%
\section{\label{s:NodelessSol}Nodeless solutions of the DNLS equations in the presence of a Scarf II potential}

In analogy with what we has been done for the NLS equation,  we can show that the solutions of the DNLS equation and the cDNLS equation in the presence of the confining Scarff-II potential are similar to the \emph{static} solutions of the equations without a potential, however the ``width'' parameter of the solution $\alpha= 2 \kappa \sqrt \omega $ is now not arbitrary but has the value $\alpha=1$, which determines the energy of the nodeless trapped solution.  The Scarff-II complex potential is $ [\, V_0 \sech^2 x + \rmi \, W_0 \sech x \tanh x \,]$.  This potential is derivable from a $\PT$-symmetric superpotential $\tilde{W}$, with $V(x) =\tilde{W}^2 \pm \tilde{W}'$, where $\tilde{W} = (m-1/2) \tanh x - i b \sech x$, and has the property that one of the partner potentials is real while the other is complex.  The derivative nonlinear \Schrodinger\ equation \cite{doi:10.1063/1.523737}
in the presence of the $\PT$-symmetric Scarff-II potential \cite{Ahmed2001343} is 
\begin{equation}\label{DN.e:1}
\fl
   \rmi \, \psi_t 
   + 
   \psi_{xx} 
   + 
   \rmi \, g( \, |\psi|^{2\kappa} \psi \, )_x 
   + 
   [\, V_0 \sech^2 x + \rmi \, W_0 \sech x \tanh x \,] \, \psi(x)  
   =
   0 \>.
\end{equation}
When $\kappa = 1$, this equation has been studied by Chen and Yan \cite{PhysRevE.95.012205}. 
This equation has been used to describe nonlinear waves in plasmas as well as optical fibers.  
In \cite{PhysRevE.95.012205}, the regimes where the trapped solutions were stable were found numerically and  mapped out as a function of $(V_0,W_0)$.  The approach we are advocating here  to study stability at arbitrary $\kappa$  is to use Derrick's theorem to understand stability when $W_0=0$, and also use the stability properties of a collective coordinate approximation to map out the approximate domain of stability.  This worked out quite well for the NLS equation when compared with a numerical study. For a particular relationship between $V_0$ and $W_0$ this potential is also derivable from a superpotential $\tilde{W}$, with $V(x) =\tilde{W}^2 \pm \tilde{W}'$.  

It is easily shown that the trapped solution of \ef{DN.e:1} for the DNLS equation in the presence Scarff-II potential is given by
\begin{subeqnarray}\label{SO.e:1}за
   \psi^{(\pm)}(x,t)
   &=
   A_0 \, \sech^{\gamma/2}(x) \, \rme^{\rmi \, [\, \omega t + \chi(x) \,]} 
   \qc
   \omega
   =
   \gamma^2/4 \>,
   \label{SO.e:1-a} \\
   g \, A_0^{2/\gamma}
   &=
   -\frac{2 W_0}{\gamma(\gamma+2)} 
   \pm
   (\gamma+1) \,
   \sqrt{1 - \frac{4 V_0}{\gamma(\gamma+2)} 
           + \Bigl [\, \frac{2 W_0}{\gamma(\gamma+2)} \,\Bigr ]^2 } \>,
   \label{SO.e:1-b} \\
   \chi(x)& = - \alpha' \tan^{-1} [\tanh(x/2)] \\
   \alpha'
   &=
   -\frac{2 W_0}{\gamma}
   \pm
   (\, \gamma + 2 \,) \, 
   \sqrt{1 -\frac{4 V_0}{\gamma(\gamma+2)} 
           + \Bigl [\, \frac{2 W_0}{\gamma(\gamma+2)} \,\Bigr ]^2 } \>,
   \label{SO.e:1-c}
\end{subeqnarray}
where $\gamma = 1 / \kappa$.
Here $\chi(x)$ is again given by 
\be
   \chi(x) = - \alpha' \tan^{-1} [\tanh(x/2)]
\ee
 We will only consider
the $(+)$ solutions in what follows since they smoothly connect with the static solutions  of the DNLS with $\beta =1$.  If instead we looked at the solutions of the cDNLS equation at arbitrary $\kappa$, then the solutions have the same form except now we obtain for the $(+)$ solution:
\begin{equation}
   \alpha'
   =
    \gamma \sqrt{
    1
      - 
      \frac{4  V_0} { \gamma (\gamma +2) }
      +
      \frac{4 W_0^2}{ \gamma^2 (\gamma + 2)^2} 
         } 
   +
   \frac{2 W_0}{ \gamma(\gamma +2) }
\end{equation}
and 
\begin{equation}
   g \, A_0^{2/\gamma}
   =
   \frac{2 W_0}{\gamma(\gamma+2)} 
   \pm
   (\gamma+1) \,
   \sqrt{1 - \frac{4 V_0}{\gamma(\gamma+2)} 
           + \Bigl [\, \frac{2 W_0}{\gamma(\gamma+2)} \,\Bigr ]^2 } \>. 
  \end{equation}
We notice that although the two $\alpha'$ are again proportional and the amplitude
the square root terms are identical, but the term outside the square root has has a positive $W_0$ term for the cDNLS solution.  For the case when $W_0=0$, which we will concentrate on in what follows, the solutions simplify and we have for the DNLS equation
\begin{subeqnarray}
   g \, A^{2/\gamma}
   &=  
   (\gamma+1) \, \sqrt{1 - \frac{4 V_0}{\gamma(\gamma+2)} } \>,
   \\
   \alpha'_1
   &=
   (\, \gamma + 2 \,) \, 
   \sqrt{1 - \frac{4 V_0}{\gamma(\gamma+2)}} \>.
\end{subeqnarray}
For the cDNLS equation we get the \emph{same} equation for the amplitude, but we find for the value of $\alpha'$ instead
\begin{equation}
   \alpha'_2
   =
   \gamma \,
   \sqrt{1 - \frac{4 V_0}{\gamma(\gamma+2)}} \>.
\end{equation}

%
%
\section{\label{s:Dissipation} Variational Principle for the DNLS equations in the presence of complex potential} 

We are interested in the stability properties of the trapped solutions of the DNLS equation in the presence of a complex trapping potential.  In our work on the NLS equation \cite{1751-8121-50-48-485205} we showed that a very useful approach to understanding stability was to study the equations for 
collective coordinates related to the lower order moment equations. These are most easily derived when the nonlinear equations can be obtained from a variational principle.  In \cite{1751-8121-50-48-485205} we used a dissipation function formalism to discuss the properties of the NLS equation in the presence of complex potentials.   What is  strikingly different for the DNLS equation is that although the DNLS equation in the absence of $W_0$ is formally a Hamiltonian dynamical system, when we formulate the equations using a variational principle, the derivative nonlinear term acts as a Raleigh dissipation term.  That is our primary reason for first  studying  the stability properties of solutions of the cDNLS equation which does not require a Dissipation Functional when $W_0$ is zero, and leads to a Hamiltonian description that is transparent. 

The DNLS equation and the cDNLS equation in the presence of a complex potential  can be obtained from a generalized Euler-Lagrange equation of the form
\begin{equation}\label{e:euler-lagrange}
   \frac{\delta \Gamma}{\delta \psi^{\ast}}
   =
   \frac{\delta \mc{F}}{\delta \psi_t^{\ast}} \>. 
\end{equation}
The action $\Gamma$ and dissipation functional $\calF$ for the DNLS equation are given by
\begin{subeqnarray}\label{e:defGF}
   \Gamma
   &=
   \tint \dd{t}
   \Bigl \{ \, 
      \frac{\rmi}{2} \tint \dd{x}
      [\, \psi^{\ast} \psi_t - \psi \psi_t^{\ast} \,]
      -
      H\,
   \Bigr \}  \equiv \int~dt (L_0- H) \>,
   \label{e:defGF-a} \\
   \mc{F}
   &=
   \tint \dd{t} F 
   =
   \tint \dd{t} [\, F_1 + F_2 \,] \>,
   \label{e:defGF-b} \\
   H
   &=
   \tint \dd{x}
   \bigl [ \,
      | \psi_x |^2
      +
      V(x) \, | \psi |^2 
   \bigr ]
   \equiv 
   H_1 - H_3 \>,
   \label{e:defGF-c} \\
   F_1
   &= 
   \rmi \tint \dd{x} 
   W(x) \, [\, \psi^\ast \psi_t - \psi^\ast_t \psi \, ] \>,
   \label{e:defGF-d} \\
   F_2
   &
   =
   \rmi g   
   \tint \dd{x} 
   \bigl \{\,
      \partial_x  \bigl [\, |\psi|^{2\kappa} \psi^\ast \, \bigr ] \, \psi_t   
      -    
      \partial_x \bigl [\, \psi|^{2\kappa}  \psi \, \bigr ] \, \psi^\ast_t \,
   \bigr \} \>.
   \label{e:defGF-e}
\end{subeqnarray}
On the other hand, for the cDNLSE, $W_0$ only enters into $\calF$, which just contains $F_1$.  For $\Gamma$ we now obtain
\begin{subeqnarray}\label{hc}
\fl
   \Gamma_c
   &= 
   \tint \dd{t} [\, L_0 - H_c \,] \>,\\
\fl
   H_c 
   &= 
   \tint \dd{x} \,
   \Bigl \{\,
      |\psi_x(x,t) |^2
      + 
      \frac{\rmi \, g \, | \psi(x,t) |^{2\kappa}}{2(\kappa+1)} \,
      [\, \psi^{\ast}(x,t) \, \psi_x(x,t) - \psi^{\ast}_x(x,t) \, \psi(x,t) \, ] \,
   \Bigr \} \>.
\end{subeqnarray}
There are three conservation laws for the cDNLS equation which are the conservation of $H$ as given by \ef{hc} as well as mass $M$ and momentum $P$ given by \ef{MP}.
In what follows we will use the fact that the Hamiltonian of the cDNLS equation when $W_0=0$ is conserved to make some general statements about the stability of the trapped solutions using Derrick's theorem.  When $W \neq 0$, one can use a collective coordinate approach based on the Raleigh Dissipation functional discussed in the appendix to study the stability of the more general solutions of both the DNLS and cDNLS equation.  For the DNLS equation the dissipation function approach using collective coordinates can be used to study the stability of the soltuions for both $V=0$ and $V \neq 0$. 
That will be left to a future paper. 

%
%
\subsection{\label{ss:gvDNLSE-Derrick}Derrick's theorem for the cDNLSE equation when $W_0=0$ }

\begin{figure}[t]
  \centering
  \includegraphics[width=0.85\columnwidth]{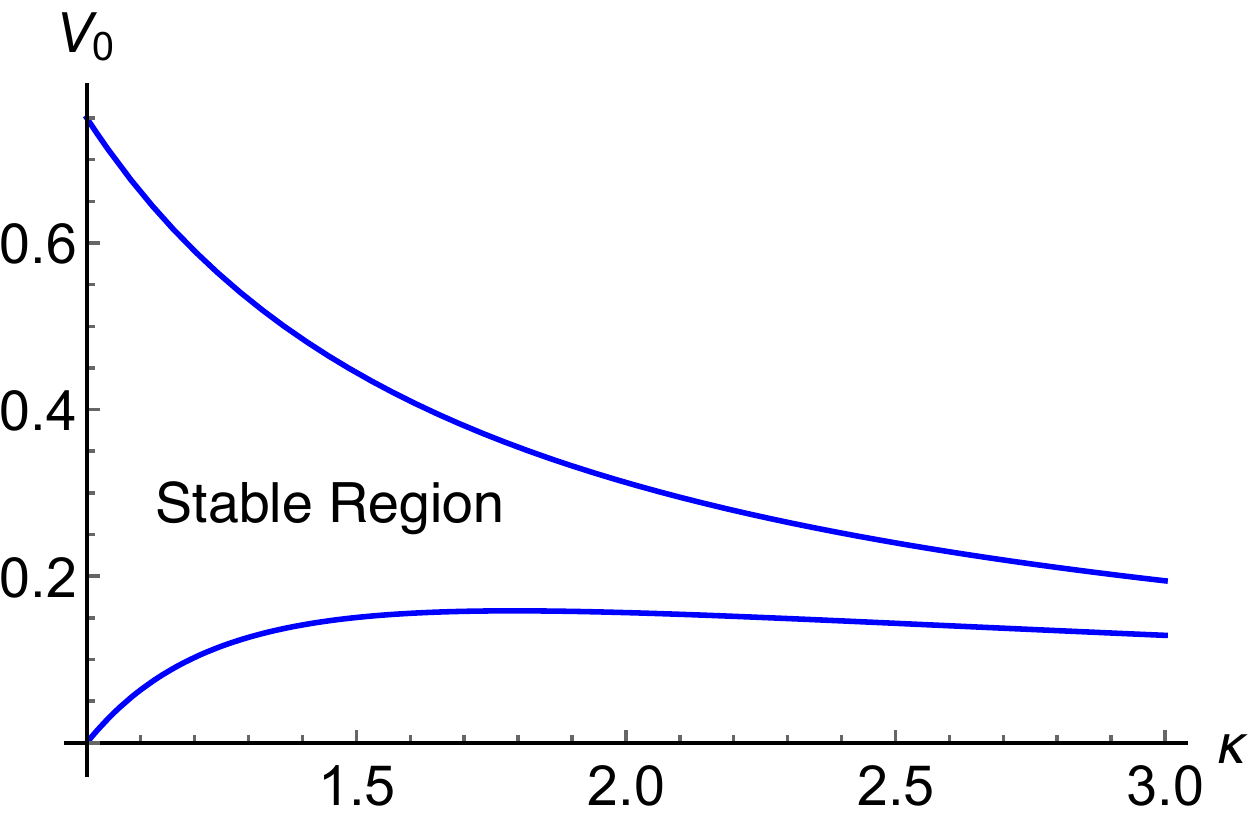}
  \caption{\label{f:crit2} The increased region of stability for $\kappa > 1$ when $V_0$ is present. Upper curve is the maximum vale of $V_0$ for a nodeless solution to exist.}
\end{figure}

When the external potential is \emph{real}, the cDNLS equation is a Hamiltonian dynamical system and one can apply Derrick's theorem to see if the solutions are stable to particular scale transformations.
Derrick's theorem \cite{doi:10.1063/1.1704233} states that an exact solution is \emph{unstable}, if under a scale transformation, $x \mapsto \eta \, x$ with the mass held fixed, the energy decreases.  To use this theorem, the energy functional as a function of $\eta$ has to be an extremum for the exact solution.  To study Derrick's theorem, we put
\begin{equation}\label{D.e:1}
   \psi_{\eta}(x)
   =
   \sqrt{\eta} \, \psi(\eta x) \>,
\end{equation}
where $\psi(x)$ is the exact static solution. The scaling \ef{D.e:1} conserves the mass.  
For the exact solution we have 
\begin{equation}\label{DV.e:3}
   g \, 
   \Bigl ( 
      \frac{M[\gamma,\zeta]}{c_1[\gamma]} 
   \Bigr )^{\frac{1}{\gamma}}
   =
   (\gamma + 1) \, \sqrt{1 - \zeta} \>,
   \qquad
   \zeta = \frac{4 V_0} {\gamma(\gamma+2)} \>.
\end{equation}
The Hamiltonians $H_1$ and $H_2$ scale in the following ways:
\begin{equation}\label{D.e:2}
   H_1[\eta,\gamma,\zeta]
   =
   \eta^2 \, h_1[\gamma,\zeta] \>,
   \qquad
   H_2[\eta,\gamma,\zeta]
   =
   \eta^{\kappa+1} \, h_2[\gamma,\zeta] \>,
\end{equation}
and are explicitly given by
\begin{eqnarray}\label{DV.e:4}
   h_1[\gamma,\zeta]&= \frac{M[\gamma,\eta]}{4} \, \gamma^2 \, 
   \Bigl ( 1- \frac{\gamma}{\gamma+1} \, \zeta \Bigr ) \>.
   \notag     \\
    h_2[\gamma,\zeta]
   &= \frac{M[\gamma,\zeta]}{2} \, \frac{\gamma^3}{\gamma+1} \, 
   ( 1 - \zeta ) \>.
   \notag
\end{eqnarray}
Note that for $\zeta = 0$, $h_2[\gamma,0] = 2 \, h_1[\gamma,0] / (\gamma +1)$,
$H_3$ does not have any simple scaling properties.  We find
\begin{eqnarray}
   H_3[\eta,\gamma,\zeta]
   &=  
   \frac{M[\gamma,\eta]}{c_1[\gamma]} \, \zeta \, \frac{\gamma (\gamma + 2)}{4}   
   g[\eta,\gamma] \>,
   \notag \\
   g[\eta,\gamma]
   &=
   \tint \dd{z} \sech^2(z/\eta) \, \sech^{\gamma}(z) \>.
   \label{DV.e:6}
\end{eqnarray}
So the scaled Hamiltonian is
\begin{eqnarray}\label{DV.e:7}
\fl
   h[\eta,\gamma,\zeta]
   &=
   \frac{H[\eta,\gamma,\zeta]}{M[\gamma,\eta]}
   \\
\fl
   &=
   \eta^2 \, \frac{\gamma^2}{4} \,
   \Bigl ( 1- \frac{\gamma}{\gamma+1} \, \zeta \Bigr )
   -
   \eta^{\kappa+1} \, \frac{\gamma^3}{2(\gamma+1)} \, 
   ( 1 - \zeta )
   -
   \zeta \, \frac{\gamma (\gamma + 2)}{4} \, 
   \frac{g[\eta,\gamma]}{c_1[\gamma]} \>.   
   \notag
\end{eqnarray}
After a bit of algebra we find that indeed 
\begin{equation}\label{DV.e:9}
   \pdv{h[\eta,\gamma,\zeta]}{\eta} \Big |_{\eta=1}
   = 
   0 \>.
\end{equation}
and that
\begin{equation}\label{DV.e:13}
   \frac{\partial^2 h[\eta,\gamma,\zeta]}{\partial \eta^2} \Big |_{\eta=1}
   =
   A[\gamma] + \zeta \, B[\gamma] > 0 \>,
\end{equation}
where 
\begin{subeqnarray}\label{DV.e:12}
\fl
   A[\gamma]
   &=
   \frac{\gamma \, (\gamma-1)}{2} \>,
   \label{DV.e:12-a} \\
\fl
   B[\gamma]
   &=
   \frac{\gamma}{2 \, (\gamma+1)(\gamma+3)} \,
   \Bigl \{\,
      \gamma^2 \, (\gamma+2) \,
      \frac{c_2[\gamma]}{c_1[\gamma]}
      -
      (\gamma - 1)(\gamma^2 + \gamma - 1) \,
   \Bigr \} \>,
   \label{DV.e:12-b}
\end{subeqnarray}
The region that is stable to changes in the width is determined by 
\begin{equation}
   \frac{\rm{d}^2 h}{\rm{d} \eta^2} \Big |_{\eta=1} > 0. 
\end{equation}
For the case $V_0=0$ one has that 
\begin{equation}
   \frac{\rm{d}^2 h}{\rm{d} \eta^2} \Big |_{\eta=1} 
   = 
   2 \, (1- \kappa) h_1 \>,
   \qquad 
   h_1 >0
\end{equation}
so that we see that when $\kappa > 1$ the solutions are unstable.  However Derrick's theorem says nothing about what happens at $\kappa=1$.  That case required the orbital stability analysis.  The same type of analysis in the NLS equation led to $\kappa=2$ being the critical value when $V=0$. 
When $V_0 > 0$ then the addition of the attractive potential increases the domain of stability beyond $\kappa=1$.  The critical regime is when
\begin{equation}
   \frac{\rm{d}^2 h}{\rm{d} \eta^2} \Big |_{\eta=1} = 0 \>, 
\end{equation}
which leads to
\begin{equation}\label{DV.e:14}
   \zeta_{\mathrm{c}}[\gamma]
   =
   - \frac{A[\gamma]}{B[\gamma]} \>,
\end{equation}
This implicitly gives the values of the  strength of the confining potential $V_0$ for a given $\kappa$ above which the solutions are stable to a change in the width parameter $\beta$ from its value of $1$. Since $0 \leq \zeta \leq 1$ there is a also a maximum value for $V_0$ as a function of $\kappa$ for a solution to exist.  When $V_0 >0$ the width stability regime is increased and we get the result shown in Fig.~\ref{f:crit2}.  Here $V_{\mathrm{max}} = \gamma(\gamma+2)/4$.
%
%
%
%

%
%
\section{\label{s:vgcdnlse} Collective Coordinate equations when $W=0$. }

In this section we will study the stability properties of the nodeless solutions to the cDNLS in the real trapping potential $V(x)$ using collective coordinate.  Here $V(x)$ is an attractive Scarff-II potential:
\begin{equation}\label{E.e:2}
   V(x)
   =
   - V_0 \, \sech^2(x) \>,
   \qquad
   V_0 > 0 \>.
\end{equation}
One of the things we would like to demonstrate is that an analysis of the linear stability of the collective coordinate approach gives essentially the same domain of stability as found using Derrick's theorem. From \ef{SO.e:1} we have when $W=0$ the nodeless trapped solutions are given by
\begin{subeqnarray}\label{E.e:3}
   \psi(x,t) 
   &=
   A \, \sech^{\gamma/2}(x) \, \rme^{ \rmi \, [\, \omega \, t  + \, \chi(x) \,] } \>,
   \label{E.e:3-a} \\
   \chi(x)
   &=
  -\alpha'  \tan^{-1}[\,\tanh(x/2)\,] \>,
   \label{E.e:3-b} 
\end{subeqnarray}
where
\begin{equation}\label{E.e:3.1}
   g \, A^{2/\gamma}
   =
   ( \gamma + 1) \, \sqrt{1-\zeta} \>,
   \qquad
   \alpha'
   =
   \gamma \, \sqrt{1-\zeta} \>,
   \qquad
   \omega
   =
    \gamma^2/4 \>,  
\end{equation}
and
\begin{equation}\label{E.e:3.2}
   \zeta
   =
   \frac{4 V_0}{\gamma(\gamma+2)}  \>,
   \end{equation}
In what follows we will suppress the prime in $\alpha'$.
The action for the cDNLS equation in the presence of a real potential is given by Eq. (\ref{hc}) , where
\be
  H_3[\psi,\psi^{\ast}]
 =
  V_0 \tint \dd{x} \sech^2(x) \, | \psi(x,t) |^2 \>.
\ee
The method of collective coordinates is described in detail in appendix B.  Our approximation is to let the 
parameters of the exact static solitary wave or the exact trapped solution become functions of time.  We introduce time dependent  coordinates related to the normalization and its conjugate as well as the first and second moments of the wave function.
That is we  consider a trial wave function given by:
\begin{subeqnarray}\label{E.e:13}
   v(x,Q(t))
   &=
   A(t) \, \sech^{\gamma/2}[\, \beta(t) y(x,t) \,] \,
   \rme^{\rmi \tchi(x,Q(t)) } \>,
   \label{E.e:13-a} \\
   \tchi(x,Q(t))
   &=
   -
   \theta(t)
   +
   p(t) \, y(x,t)
   +
   \Lambda(t) \, y^2(x,t)
   -
    \chi(\beta(t) y(x,t)) \>,
   \label{E.e:13-b}
\end{subeqnarray}
where $y(x,t) = x - q(t)$ and $\chi(x)$ is given in \ef{E.e:3-b}.  The collective parameters here are:
\begin{equation}\label{E.e:14}
   Q(t)
   =
   \{\,\theta(t),A(t),q(t),p(t),\beta(t),\Lambda(t) \,\} \>.
\end{equation}
If we were to include $W(x)$ we would need all six variational parameters to study the stability problem, since then the mass $M$ would be come a function of time and this variable is conjugate to the phase $\theta(t)$.  When $M$ is conserved $\theta$  does not enter into the dynamics and one only needs four collective coordinates to capture the dynamics. 
For the NLS equation we have used this approximation to study the behavior of exact solitary wave solutions as they pass through a complex Scarff-II potential \cite{PhysRevE.94.032213}, as well as to study the stability of the trapped solutions \cite{1751-8121-50-48-485205}.  Here we are concerned with the stability of the trapped nodeless solutions as a function of the parameter $\kappa$.  For that case we assume that the mass of the wave is conserved and equal to the mass of the exact solution.  To study the stability we will allow the width and the position to be slightly perturbed initially. 
That is we will choose 
\begin{eqnarray}\label{E.e:15}
   q_0
   =
   \delta q_0 \>,
   \quad
   p_0
   =
   0 \>,
   \quad
   \beta_0
   =
   1+ \delta \beta_0 \>,
   \quad
   \Lambda_0
   =
   0 \>,
   \quad
   \dot{\theta}_0
   =
   - \beta_0^2 \gamma^2/4 \>,
   \notag
\end{eqnarray}
For the problem of a traveling solitary wave encountering an external potential we would instead choose $\dot q_0 \neq 0$ and start the solitary wave outside the range of the external potential.  We would also use the exact moving solution as the basis for our variational approximation.

When $W_0=0$, the normalization $M$ is conserved so that
\bea \label{E.e:16}
  M_s&&
   \equiv
   \tint \dd{x} |v(x,Q)|^2
   =
   \frac{A^2(t)}{\beta(t)} \tint \dd{z} \sech^{\gamma}(z)
   =
   \frac{A^2(t) \, c_1[\gamma]}{\beta(t)} \>, \nonumber \\
  &&  =
   \Bigl [\,
      \frac{ ( \gamma + 1) \, \sqrt{1-\zeta}}{g} \, 
   \Bigr ]^{\gamma} \, 
c_1[\gamma] \>.
\eea
so we can eliminate $A(t)$ as a dynamic variable in favor of $\beta(t)$,
\begin{equation}\label{E.e:17}
   A(t)
   =
   \sqrt{\frac{M_s \beta(t)}{c_1[\gamma]} }
   =
   \sqrt \beta 
   \Biggl [
      \frac{ ( \gamma + 1) ~ \sqrt{1-\zeta}} {g} \,
   \Biggr ]^{\frac{\gamma}{2}}
   = 
   A_0 \sqrt{\frac{\beta(t)}{\beta_0}} \>.
\end{equation}
The remaining dynamic variables are
\begin{equation}\label{E.e:18}
   Q(t)
   =
   \{\,q(t),p(t),\beta(t),\Lambda(t) \,\} \>.
\end{equation}
The trial wave function  can be written as:
\begin{equation}\label{E.e:19}
   v(x,Q)
   =
   A_0 \sqrt{\beta(t)/\beta_0} \, 
   \sech^{\gamma/2}[\, \beta(t) y(x,t) \,] \,
   \rme^{\rmi \tchi(x,Q(t)) } \>,
\end{equation}
We find that 
\begin{eqnarray}\label{E.e:26}
\fl
   L_0[Q,\dot{Q}]
   &=
   \frac{\rmi}{2}
   \tint \dd{x} [\, v^{\ast} v_x - v^{\ast}_x v \,]
   \\
   \fl
   &=
   - A_0^2 \, (\beta/\beta_0) \,
   \tint \dd{y} \sech^{\gamma}( \beta y ) \,
   [\, 
   -
   p \, \dot{q}
   +
   \dot{\Lambda} \, y^2
   +
   (\alpha/2) \beta \, \dot{q} \, \sech(\beta y) \,
   ] 
   \notag \\
   \fl
   &=
   M_s \, 
   \Biggl \{\,
      \dot{\theta}
      +
      \Bigl [\,
         p - \alpha \beta \frac{c_1[\gamma+1]}{2 \,c_1[\gamma]}\,
      \Bigr ] \, \dot{q}
      -
      \frac{\dot{\Lambda}}{\beta^2} \, \frac{c_2[\gamma]}{c_1[\gamma]} \,
   \biggr \}
   =
   \pi_{\mu}[Q] \, \dot{Q}^{\mu} \>.
   \notag
\end{eqnarray}
So the only non-zero derivatives of the $\pi_\mu$ are 
\begin{equation}\label{E.e:28}
   \partial_p \pi_q 
   = 
   M_s \>,
   \quad
   \partial_\beta \pi_q
   =
   - M_s \, \alpha \, \frac{c_1[\gamma+1]}{2\,c_1[\gamma]} \>,
   \quad
   \partial_\beta  \pi_\Lambda 
   =
   \frac{ 2 M_s \,c_2[\gamma] }{ \beta^3 \,  c_1[\gamma] } \>.
\end{equation}
Following the general framework laid out in Appendix B, the antisymmetric symplectic tensor $ f_{\mu\nu}[Q]$ is then given by
\begin{equation}\label{E.e:29}
   f_{\mu\nu}[Q]
   =
   M_s
   \left (
   \begin{array}{cccc}
      0 & -1 & D & 0 \\
      1 & 0 & 0 & 0 \\
     -D & 0 & 0 & C \\
      0 & 0 & - C & 0
   \end{array}
   \right ) \>,
\end{equation}
where
\begin{equation}\label{E.e:30}
   C
   =
   \frac{ 2 \,c_2[\gamma] }{ \beta^3 \,  c_1[\gamma] } \>,
   \quad
   D
   =
   \alpha \, \frac{c_1[\gamma+1]}{2\,c_1[\gamma]} \>.
\end{equation}
The determinant is then given by
\begin{equation}\label{E.e:31}
   \det\{f_{\mu\nu}[Q]\}
   =
   M_s^4 C^2 \>,
\end{equation}
and the inverse by
\begin{equation}\label{E.e:32}
   f^{\mu\nu}[Q]
   =
   \frac{1}{M_s}
   \left (
   \begin{array}{cccc}
      0 & 1 & 0 & 0 \\
      -1 & 0 & 0 & -D/C \\
      0 & 0 & 0 & -1/C \\
      0 & D/C & 1/C & 0
   \end{array}
   \right ) \>.
\end{equation}
For the Hamiltonian, we find 
\begin{eqnarray}\label{E.e:33}
\fl
   H_1[Q]
   &=
   \tint \dd{x} | v_x(x,Q) |^2
   \\
   \fl
   &=
   M_s \,
   \Bigl \{\,
      \frac{\beta^2}{4} \, \frac{\gamma^2}{\gamma+1}
      +
      p^2
      +
      \frac{ 4 \, \Lambda^2}{\beta^2} \, \frac{c_2[\gamma]}{c_1[\gamma]}
      +
      \frac{\alpha^2 \beta^2}{4} \, \frac{\gamma}{\gamma+1}
      -
      \alpha \, \beta \, p \, \frac{c_1[\gamma+1]}{c_1[\gamma]} \,
   \Bigr \}  \>.
   \notag
\end{eqnarray}

\begin{eqnarray}\label{E.e:34}
\fl
   H_2[Q]
   &=
   \frac{g}{\kappa+1}
   \tint \dd{x}
   | v(x,Q) |^{2\kappa} \,
   \frac{\rmi}{2} \, 
   [\, v^{\ast}(x,Q) \, v_x(x,Q) - v^{\ast}_x(x,Q) \, v(x,Q) \, ] \>.
   \\
   \fl
   &=
   - 
   g \, M_s \,
   \Bigl( \frac{M_s \, \beta}{c_1[\gamma]} \Bigr )^{\frac{1}{\gamma}} \,
   \Bigl \{\,
      p \, \frac{\gamma}{\gamma+1} \, 
      \frac{c_1[\gamma+1]}{c_1[\gamma]}
      -
      \beta \, \frac{\alpha \, \gamma^2}{2\,(\gamma+1)^2} \, 
   \Bigr \}
   \notag \\
   \fl
   &=
   -
   M_s \, \, (\gamma+1) \, \sqrt{1-\zeta} \,
 \beta^{\frac{1}{\gamma}} \ \,
   \Bigl \{\,
      p \, \frac{\gamma}{\gamma+1} \, 
      \frac{c_1[\gamma+1]}{c_1[\gamma]}
      -
      \beta \, \frac{\alpha \, \gamma^2}{2\,(\gamma+1)^2} \, 
   \Bigr \} \>.
   \notag    
\end{eqnarray}
\begin{eqnarray}\label{E.e:34.1}
\fl
   H_3[Q]
   &=
   V_0 \tint \dd{x} \sech^2(x) \, | \psi(x,t) |^2
   \\
   \fl
   &=
   M_s \, \zeta \,
   \frac{\gamma(\gamma+2)}{4} \, \frac{\beta}{c_1[\gamma]}
   \tint \dd{y} \sech^{\gamma}(\beta y) \sech^2(y+q)
   \notag \\
   \fl
   &=
   M_s \, \zeta \, \beta \, \frac{\gamma(\gamma+2)}{4} \,
   \frac{I_2[\gamma,\beta,q]}{c_1[\gamma]} \>.
   \notag
\end{eqnarray}
The full Hamiltonian is then given by
\begin{eqnarray}\label{E.e:35}
\fl
   H[Q]
   &=
   M_s \,
   \Bigl \{\,
      p^2
      +
      \frac{\beta^2}{4} \, \frac{\gamma \, ( \alpha^2 + \gamma )}{\gamma+1}
      +
      \frac{ 4 \, \Lambda^2}{\beta^2} \, \frac{c_2[\gamma]}{c_1[\gamma]}
      -
      \alpha \, \beta \, p \, \frac{c_1[\gamma+1]}{c_1[\gamma]}
      \\
      \fl
      & \quad
      +
      \, \sqrt{1-\zeta} \,
    \beta^{\frac{1}{\gamma}} \ \,
      \Bigl [\,
         p \, \gamma \, 
         \frac{c_1[\gamma+1]}{c_1[\gamma]}
         -
        \beta \, \frac{\alpha \, \gamma^2}{2\,(\gamma+1)} \, 
      \Bigr ]
      -
      \zeta \, \beta \, \frac{\gamma(\gamma+2)}{4} \,
      \frac{I_2[\gamma,\beta,q]}{c_1[\gamma]} \,
   \Bigr \} \>.
   \notag
\end{eqnarray}
For the symplectic formalism, we will need the derivatives $u_{\mu} = \partial_{\mu} H[Q]$.  Using results in \ref{s:Integrals}, we find
\begin{subeqnarray}\label{E.e:36}
\fl
   u_q
   &=
   M_s \,
   \Bigl \{\,
      \zeta \, \beta \, \frac{\gamma(\gamma+2)}{2} \,
      \frac{f_3[\gamma,\beta,q]}{c_1[\gamma]} \,
   \Bigr \} \>,
   \label{E.e:36-a} \\
\fl
   u_p
   &=
   M_s \,
   \Bigl \{\,
      2 \, p
      -
      \alpha \, \beta \, \frac{c_1[\gamma+1]}{c_1[\gamma]}
      +
      \, \sqrt{1-\zeta} \, \gamma \,
    \beta^{\frac{1}{\gamma}} \ \,
      \frac{c_1[\gamma+1]}{c_1[\gamma]} \,
   \Bigr \} \>,
   \label{E.e:36-b} \\
\fl
   u_{\beta}
   &=
   M_s \,
   \Bigl \{\,
      \frac{\beta}{2} \, \frac{\gamma \, ( \alpha^2 + \gamma )}{\gamma+1}
      -
      \frac{ 8 \, \Lambda^2}{\beta^3} \, \frac{c_2[\gamma]}{c_1[\gamma]}
      -
      \alpha \, p \, \frac{c_1[\gamma+1]}{c_1[\gamma]}
      \notag \\
\fl
      & \qquad
      +
       \, \sqrt{1-\zeta} \,
    \beta^{\frac{1}{\gamma}} \ \,
      \Bigl [\,
         \frac{p}{\beta} \, \frac{c_1[\gamma+1]}{c_1[\gamma]}
         -
         \frac{\alpha \, \gamma}{2} \, 
      \Bigr ]
      \label{E.e:36-c} \\
\fl
      & \qquad
      -
      \zeta \, \frac{\gamma(\gamma+2)}{4} \, 
      \frac{I_2[\gamma,\beta,q]}{c_1[\gamma]}
      +
      \zeta \, \beta \, \frac{\gamma^2(\gamma+2)}{4} \,
      \frac{f_6[\gamma,\beta,q]}{c_1[\gamma]} \,
   \Bigr \} \>,
   \notag \\
   \fl
   u_{\Lambda}
   &=
   M_s \,
   \Bigl \{\,
      \frac{ 8 \, \Lambda}{\beta^2} \, \frac{c_2[\gamma]}{c_1[\gamma]}
   \Bigr \} \>.
   \label{E.e:36-d}
\end{subeqnarray}
So using  \ef{E.e:32}, we find
that  the dynamic equations are:
\begin{subeqnarray}\label{E.e:41}
\fl
   \dot{q}
   &=
   2 \, p
   -
   \Bigl [\,
      \alpha \, \beta 
      -
   \, \gamma \, \sqrt{1-\zeta} \,
    \beta^{\frac{1}{\gamma}} \ \,
   \Bigr ]\,
   C_2[\gamma] \>,
   \label{E.e:41-a} \\
\fl
   \dot{p}
   &=
   -
   \zeta \, \beta \, \frac{\gamma(\gamma+2)}{2} \,
   \frac{f_3[\gamma,\beta,q]}{c_1[\gamma]}
   -
   2 \, \alpha \, \beta \, \Lambda \, C_2[\gamma] \>,
   \label{E.e:41-b} \\
\fl
   \dot{\beta}
   &=
   - 4 \, \beta \,  \Lambda \>,
   \label{E.e:41-c} \\
\fl
   \dot{\Lambda}
   &=
   - 
   4 \, \Lambda^2
   +
   \frac{ \,\beta^2 \, p}{2} \, \sqrt{1-\zeta} \,
 \beta^{\frac{1}{\gamma}} \ \,
   C_3[\gamma] \>,
   \label{E.e:41-d} \\
\fl
   & \qquad
   +
   \frac{\gamma \, \beta^3}{4} \,
   \Bigl [\,
      \beta \, \frac{\alpha^2 + \gamma }{\gamma+1}
      -
      \alpha \,  \, \sqrt{1-\zeta} \,
    \beta^{\frac{1}{\gamma}} \ \,
   \Bigr ] \, \frac{c_1[\gamma]}{c_2[\gamma]}   
   \notag \\
\fl
   & \qquad
   -
   \frac{\alpha \, \beta^3}{4} \,
   \Bigl [\,
      \alpha \, \beta 
      -
       \, \gamma \,  \sqrt{1-\zeta} \,
    \beta^{\frac{1}{\gamma}} \ \,
   \Bigr ] \,
   \frac{c_1^2[\gamma+1]}{c_1[\gamma]\,c_2[\gamma]}
   \notag \\
\fl
   & \qquad
   -
   \zeta \, \frac{\beta^3 \, \gamma(\gamma+2)}{8} \, 
   \frac{I_2[\gamma,\beta,q]}{c_2[\gamma]}
   +
   \zeta \, \frac{\beta^4 \, \gamma^2(\gamma+2)}{8} \,
   \frac{f_6[\gamma,\beta,q]}{c_2[\gamma]} \>.
   \notag               
\end{subeqnarray}
Let us note here that when $\zeta = 0$ and $\alpha = \gamma$, Eqs.~\ef{E.e:41} are the appropriate collective coordinate eqations for $V_0=0$ ($\zeta=0$). 
\begin{equation}\label{E.e:42}
\fl
   \dot{\Lambda}
   =
   - 
   4 \, \Lambda^2
   +
   +
   \frac{ \, \beta^2 \, p }{2} \,
  \beta^{\frac{1}{\gamma}} \ \,
   C_3[\gamma]
   +
   \frac{\gamma^2 \, \beta^3}{4} \,
   \Bigl [\,
      \beta 
      -
       \,
    \beta^{\frac{1}{\gamma}} \ \,
   \Bigr ] \, C_1[\gamma] \>.
\end{equation}
As a further check, let us set $\alpha = \gamma \, \sqrt{1 - \zeta}$, which is the value for the solitary wave solution at $t=0$.  Then \ef{E.e:41} becomes:
\begin{subeqnarray}\label{E.e:43}
\fl
   \dot{q}
   &=
   2 \, p
   -
   \gamma \, \sqrt{1-\zeta} \,
   \Bigl [\,
      \beta 
      -
       \, 
    \beta^{\frac{1}{\gamma}} \ \,
   \Bigr ]\,
   C_2[\gamma] \>,
   \label{E.e:43-a} \\
\fl
   \dot{p}
   &=
   -
   \zeta \, \beta \, \frac{\gamma(\gamma+2)}{2} \,
   \frac{f_3[\gamma,\beta,q]}{c_1[\gamma]}
   -
   2 \, \gamma \, \beta \, \Lambda \, \sqrt{1-\zeta} \, C_2[\gamma] \>,
   \label{E.e:43-b} \\
\fl
   \dot{\beta}
   &=
   = - 4 \, \beta \,  \Lambda \>,
   \label{E.e:43-c} \\
\fl
   \dot{\Lambda}
   &=
   - 
   4 \, \Lambda^2
   +
   \frac{ \,\beta^2 \, p}{2} \, \sqrt{1-\zeta} \,
 \beta^{\frac{1}{\gamma}} \ \,
   C_3[\gamma]
   \label{E.e:43-d} \\
\fl
   & \qquad
   +
   \frac{\gamma^2 \, \beta^3}{4} \,
   \Bigl [\,
      \beta \, \frac{\gamma \, ( 1 - \zeta ) + 1 }{\gamma+1}
      -
       \, (1-\zeta) \,
    \beta^{\frac{1}{\gamma}} \ \,
   \Bigr ] \, \frac{c_1[\gamma]}{c_2[\gamma]}   
   \notag \\
\fl
   & \qquad
   -
   \frac{\gamma^2 \, \beta^3}{4} \, (1 - \zeta) \,
   \Bigl [\,
      \beta 
      -
       \, 
    \beta^{\frac{1}{\gamma}} \ \,
   \Bigr ] \,
   \frac{c_1^2[\gamma+1]}{c_1[\gamma]\,c_2[\gamma]}
   \notag \\
\fl
   & \qquad
   -
   \zeta \, \frac{\beta^3 \, \gamma(\gamma+2)}{8} \,
   \Bigl [\,
       \frac{I_2[\gamma,\beta,q]}{c_2[\gamma]}
       -
       \beta \, \gamma \, \frac{f_6[\gamma,\beta,q]}{c_2[\gamma]} \,
   \Bigr ]  \>.
   \notag               
\end{subeqnarray}
%
%
\section{\label{s:Vlinear}Linear response}

The stability of the trapped solution can be studied by looking at the small oscillations about the exact solution.  The collective coordinate equations are of the form:
\begin{equation}\label{LS.e:1}
   \dot{Q}^{\mu}(t)
   =
   F^{\mu}[Q(t)] \>.
\end{equation}
Setting $Q^{\mu}(t) = Q_0^{\mu} + \delta Q^{\mu}(t)$, and expanding \ef{LS.e:1} out to first order gives:
\begin{equation}\label{LS.e:2}
   \delta \dot{Q}^{\mu}(t)
   =
   F^{\mu}[Q_0] + M^{\mu}{}_{\nu}[Q_0] \, \delta Q^{\mu}(t) \>.
\end{equation}
From \ef{LS.e:2},
\begin{equation}\label{L2.e:5}
   \delta \ddot{Q}^{\mu}(t)
   +
   \Omega^{\mu}_{\nu}[Q_0] \, \delta Q^{\nu}(t)
   =
   R^{\mu}[Q_0] \>,
\end{equation}
where
\begin{eqnarray}\label{L2.e:6}
   \Omega^{\mu}_{\nu}[Q_0]
   &=
   - M^{\mu}{}_{\sigma}[Q_0] \, M^{\sigma}{}_{\nu}[Q_0] \>,
   \\
   R^{\mu}[Q_0]
   &=
   M^{\mu}{}_{\nu}[Q_0] \, F^{\nu}[Q_0] \>.
   \notag
\end{eqnarray}
The eigenvalues of $\Omega^{\mu}{}_{\nu}[Q_0]$ determine the square of the oscillation frequencies $\omega^2[Q_0]$ about the equilibrium values $F^{\mu}[Q_0]$.  

Expanding the dynamic equations about the solitary wave initial conditions, 
\begin{equation}\label{vL.e:1}
   q_0
   =
   0 \>,
   \quad
   p_0
   =
   0 \>,
   \quad
   \beta_0   
   =
   1 \>,
   \quad
   \Lambda_0
   =
   0 \>,
   \quad
   \alpha
   =
   \gamma \, \sqrt{1-\zeta} \>,
\end{equation}
we find  $F^{\mu}[Q_0] = 0$ and that the matrix $M$ is of the form:
\begin{equation}\label{vL.e:2}
   M^{\mu}{}_{\nu}
   =
   \left (
   \begin{array}{cccc}
      0 & A & B & 0 \\
      C & 0 & 0 & D \\
      0 & 0 & 0 & E \\
      0 & F & G & 0
   \end{array}
   \right ) \>,
\end{equation}
where
\begin{subeqnarray}\label{z.e:1}
\fl
   A
   &=
   2 \>,
   \label{z.e:1-a} \\
\fl
   B
   &=
   - \sqrt{1-\zeta} \, (\, \gamma - 1 \,) \, C_2[\gamma] \>,
   \label{z.e:1-b} \\
\fl
   C
   &=
   - \frac{ \gamma^3 \, (\gamma+2) \, \zeta }{2 \, (\gamma+1)(\gamma+3) } \>,
   \label{z.e:3-c} \\
\fl
   D
   &=
   - 2 \, \gamma \, C_2[\gamma] \>,
   \label{z.e:1-d} \\
\fl
   E
   &=
   - 4 \>,
   \label{z.e:1-e} \\
\fl   
   F
   &=
   \frac{1}{2} \, \sqrt{1-\zeta} \, C_3[\gamma] \>,
   \label{z.e:1-f} \\
\fl
   G
   &= G_1+G_2+G_3+G_4
\end{subeqnarray}
and with
\begin{subeqnarray}\label{z.e:2}
\fl
   G_1
   &=
   \frac{\gamma^3 \, (\gamma + 2) \, \zeta}{4\, (\gamma+1)(\gamma+3)} \>,
   \label{z.e:2-a} \\
\fl
   G_2
   &=
   \gamma \,
   \Bigl \{\,
      \frac{\gamma}{1 + \gamma} \, [\, 1 + \gamma \, (1 - \zeta) \, ]
      -
      \frac{1 + 3 \gamma}{4} \,  (1 - \zeta)
   \Bigr \} \>,
   \label{z.e:2-b} \\
\fl
   G_3
   &=
   -
   (1 - \zeta) \, \gamma \, 
   \Bigl [\, 
      \gamma
      -
      \frac{1 + 3 \, \gamma}{4} \,  
   \Bigr ] \>.
   \label{z.e:2-c} \\
\fl
   G_4
   &=
   -
   \frac{ \gamma \, (\, 4 + 11 \, \gamma + \gamma^2 \,) \, \zeta}
        { 4 \, (1 + \gamma) (3 + \gamma) } \>.
   \label{z.e:2-d}
\end{subeqnarray}

%
%
\subsection{\label{ss:Vcritical}Critical value of $\zeta$}

\begin{figure}[t]
   \centering
   \subfigure[\ $\zeta_{\mathrm{c}}$]
   { \label{f:fig8-a} 
   \includegraphics[width=0.45\columnwidth]{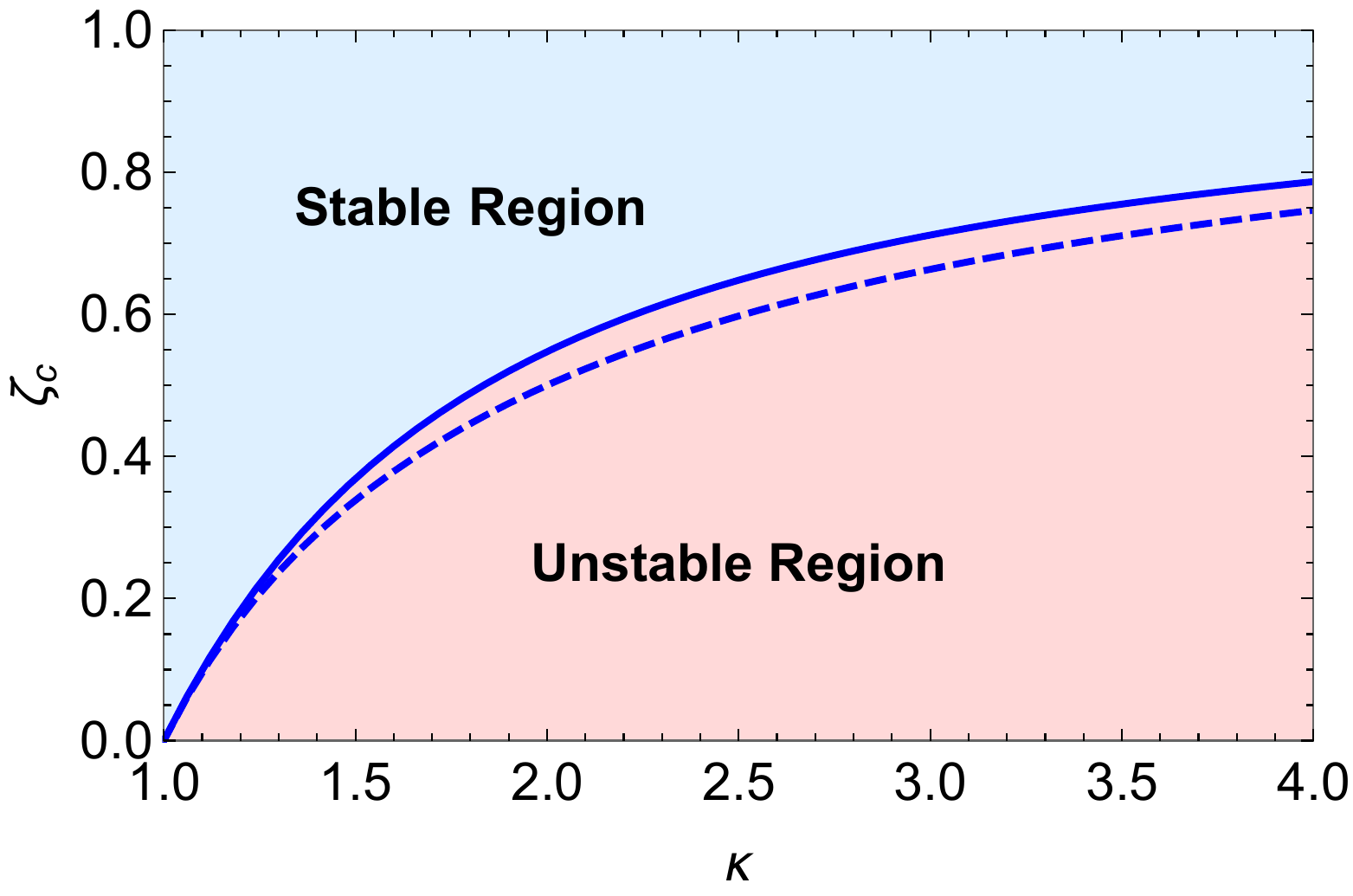} }
   \subfigure[\ $V_{\mathrm{c}}$]
   { \label{f:fig8-b} 
   \includegraphics[width=0.45\columnwidth]{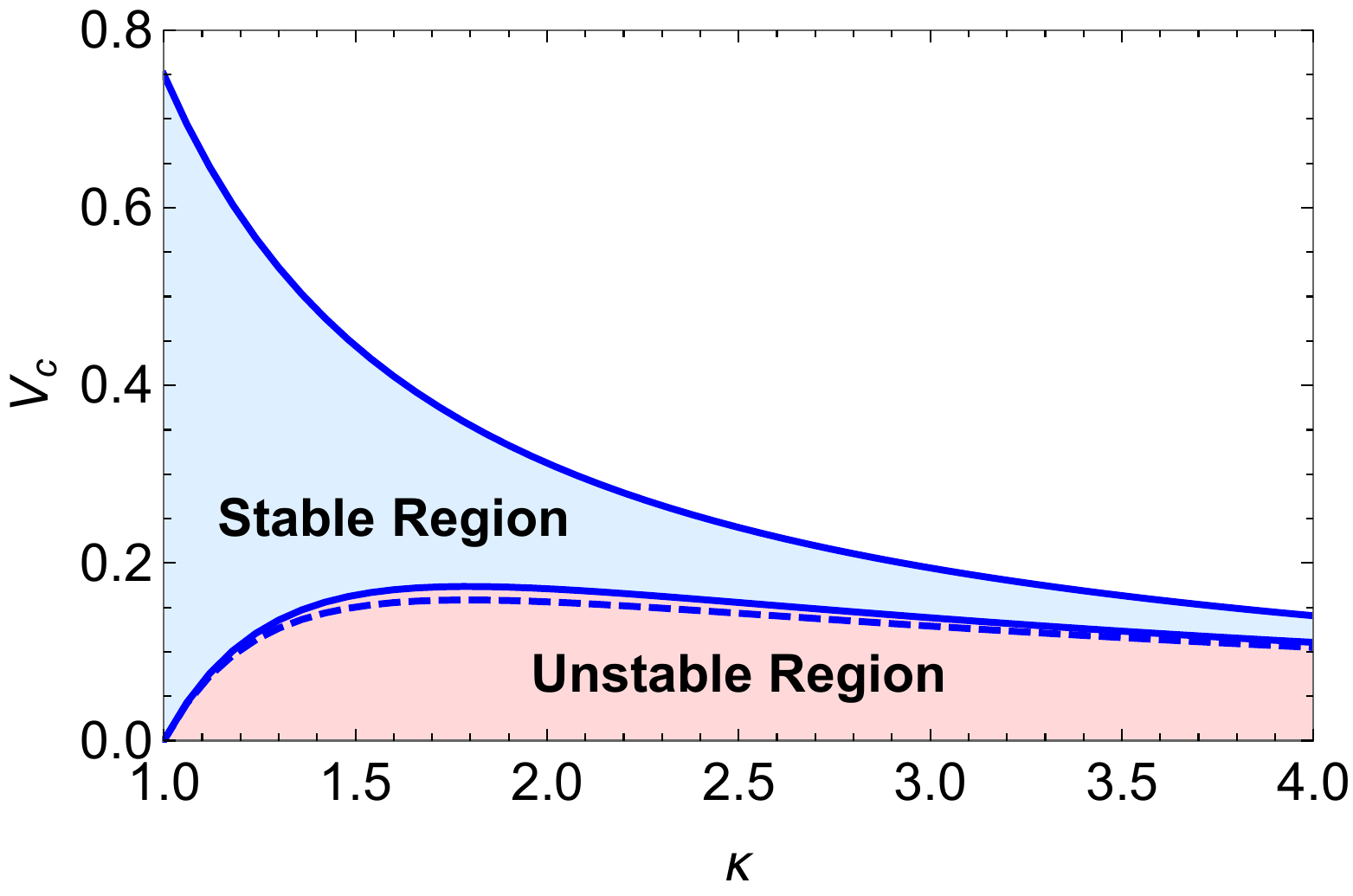} }
   \caption{\label{f:fig8}Plot of (a) $\zeta_{\mathrm{c}}$ and 
   (b) $V_{\mathrm{c}}$ as a function
   of $\kappa$ for $\xi = 1.0$.  The solid blue lines are the result
   for the four-parameter collective ansatz and the dashed blue line
   is the result of Derrick's theorem.}
\end{figure}

For the case when the initial value of the variational ansatz is the exact soliton solution so that $q_0 = p_0 = \Lambda_0 = 0$ and $\beta_0 = 1$, the critical value of $\zeta_{\mathrm{c}}[\kappa]$ is when $\omega_{\pm}^2[1,\zeta_{\mathrm{c}},\gamma] = 0$.  Numerical results are shown in Fig.~\ref{f:fig8} and compared to results found using Derrick's theorem.  The stable region is for $\zeta_{\mathrm{c}}[\gamma] \le \zeta \le 1$, or for
\begin{equation}\label{z.e:3}
   V_{\mathrm{c}}[\gamma]
   = 
   \zeta_{\mathrm{c}}[\gamma] \,  \, V_{\mathrm{max}}[\gamma] 
   \le V_0 \le V_{\mathrm{max}}[\gamma] \>.
\end{equation}
%
%
%
%

%
%
\section{\label{s:Vdynamics}Dynamics of the Collective Coordinates}

Here we solve the collective coordinate equations at $\kappa=2$ to show the  evolution of the nodeless solution once it is perturbed.
In the stable regime the collective coordinates undergo quasiperiodic motion around the exact solution whereas in the unstable regime, the solution leaves the area of the potential and also narrows greatly. Here $G(t)$ is related to the average value of the fluctuation $(x-q(t))^2$ which is the ``width'' of the solitary wave. 

\begin{figure}[t]
   \centering
   \subfigure[\ $q(t), \zeta = 0.8$]
   { \label{f:fig9-a} 
   \includegraphics[width=0.45\columnwidth]{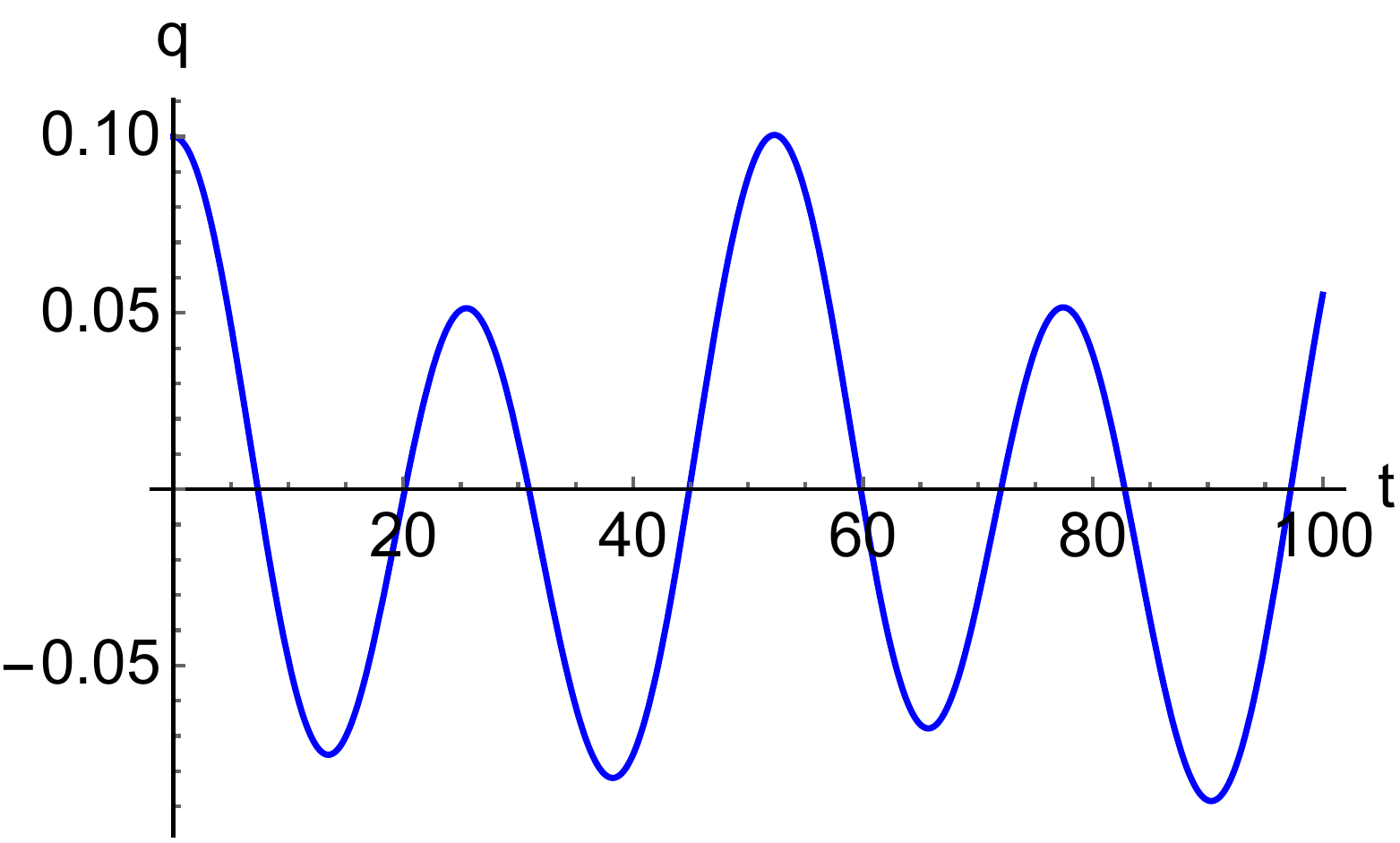} }
   \subfigure[\ $G(t), \zeta = 0.8$]
   { \label{f:fig9-b} 
   \includegraphics[width=0.45\columnwidth]{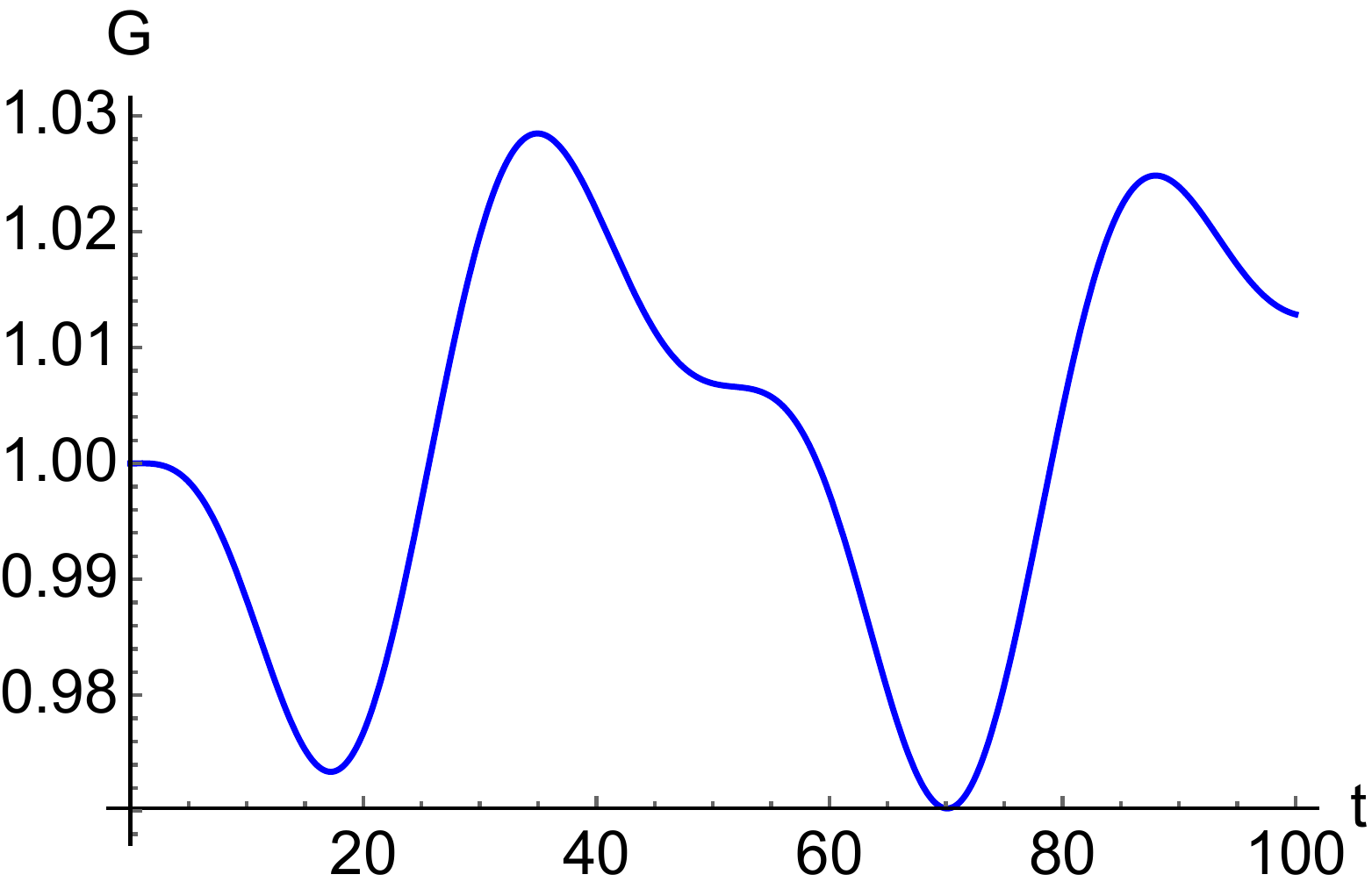} }
   \subfigure[\ $q(t), \zeta = 0.2$]
   { \label{f:fig9-c} 
   \includegraphics[width=0.45\columnwidth]{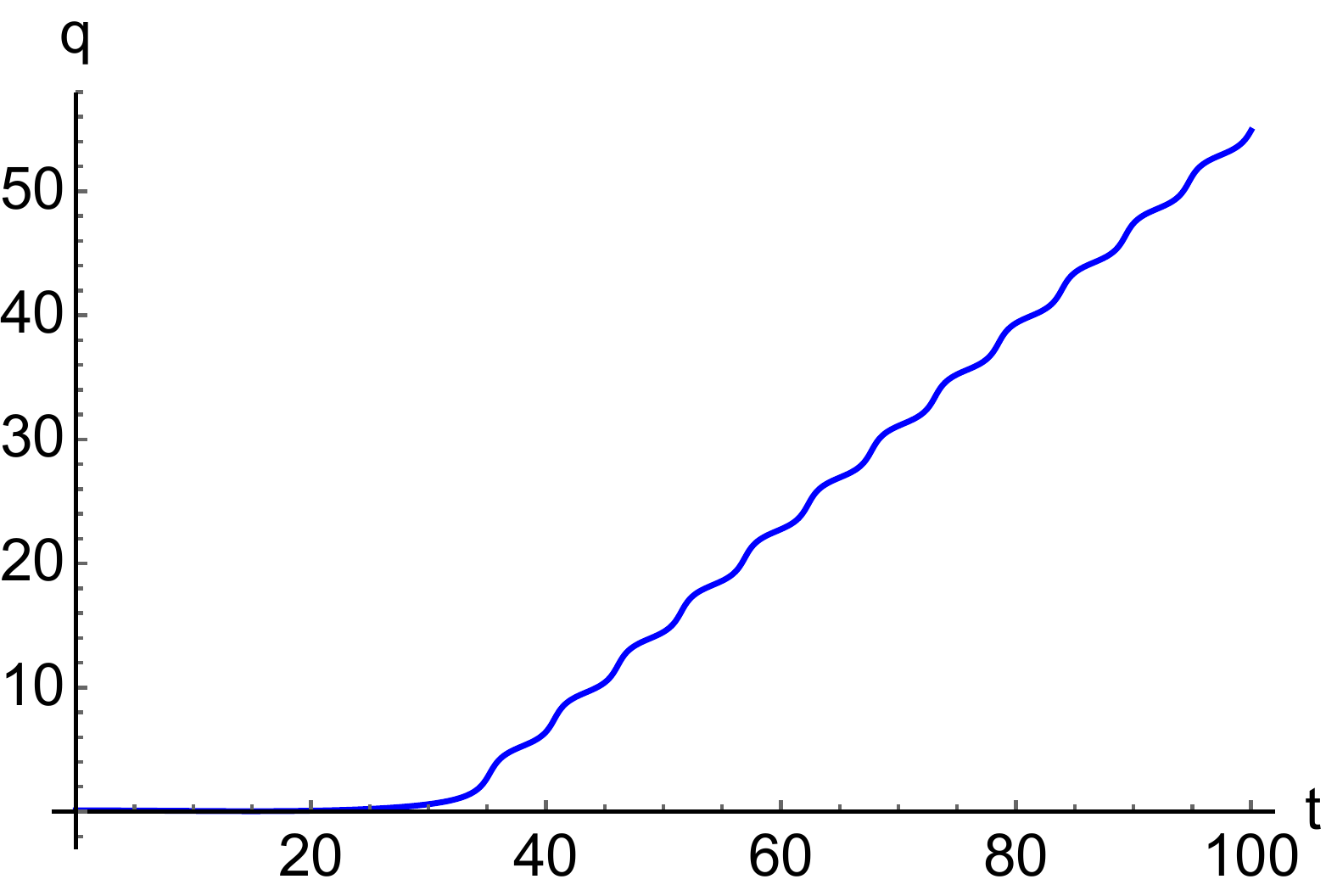} }
   \subfigure[\ $G(t), \zeta = 0.2$]
   { \label{f:fig9-d} 
   \includegraphics[width=0.45\columnwidth]{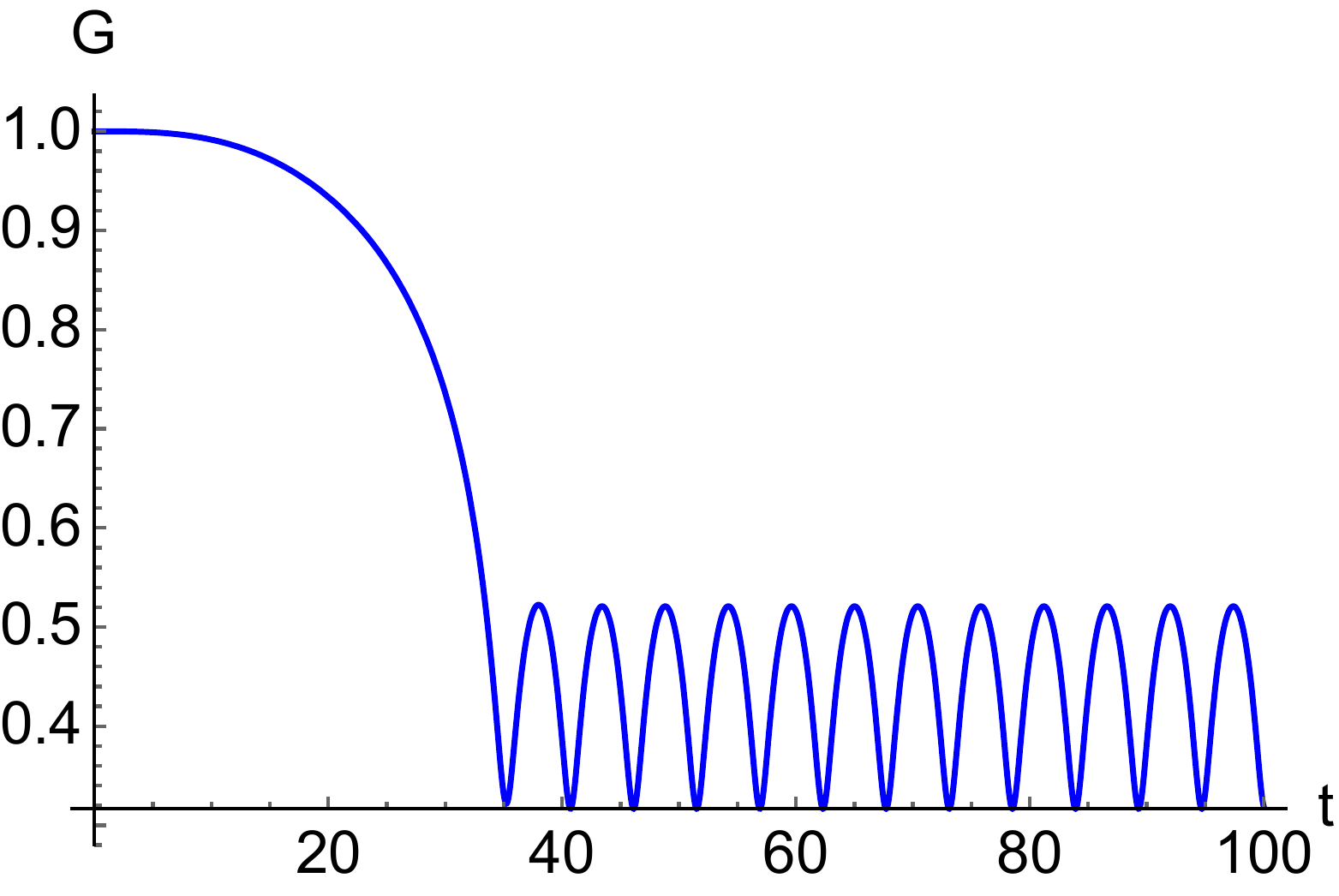} }
   \caption{\label{f:fig9}Plots of $q(t)$ and $G(t) = 1 / \beta(t)$
   for $\kappa=2$ and $\zeta=0.8$ in figures (a) and (b) (the stable
   region) and $\zeta=0.2$ in figures (c) and (d) (the unstable
   region).  Here $q_0 = 0.1$, $p_0 = 0$, $G_0 = 1$, and $\Lambda_0 = 0$.}
\end{figure}

In Fig.~\ref{f:fig9} we plot $q(t)$ and $G(t)$ from solutions of the dynamics equations \ef{E.e:43} for the case with $\kappa = 2$ and $q_0 = 0.1$, $p_0 = 0$, $G_0 = 1$, and $\Lambda_0 = 0$.   According to the analysis in Fig.~\ref{f:fig8}, the dynamic simulation in Figs.~\ref{f:fig9-a} and \ref{f:fig9-b} with $\zeta = 0.8$ should be in the stable region, whereas for Figs.~\ref{f:fig9-c} and \ref{f:fig9-d} with $\zeta = 0.2$ the dynamic simulation should be in the unstable region.  The collective coordinate dynamics seems to support this conclusion.  

%
%
%
%

%
%
\section{\label{s:Conclusions}Conclusions}

In this paper we found new solutions to the derivative nonlinear \Schrodinger\ as well as the related cDNLS equation which is a  Hamiltonian dynamical system.  For the related cDNLS equation we were able to map out the domain of orbital stability  of the solitary wave solutions as a function of $\kappa$, and $\sigma =c/(2 \sqrt \omega)$, where $c$ is the velocity and $\omega$ is the frequency of the solitary wave. When the external potential was a real Scarff-II potential, we were able to study the stability of the trapped nodeless solutions as a function of $\kappa$ and $V_0$ using Derrick's theorem and a collective coordinate approach.  The two methods gave essentially the same regimes of stability.  The results we obtained for the trapped solutions are similar to those obtained for the NLS equation in the presence of this potential, with the crucial difference that the critical value of $\kappa$ for the cDNLSE was $1$ instead of $2$ for the NLS equation.  The solitary wave solutions of these two equations are quite different.  Solutions to the cDNLSE solutions only exist when $\sigma <1$ and their analytic form for $c>0$ is quite different from the solitary wave solutions of the NLSE. When the potential is complex, we showed in the appendix that one can obtain the equations from a dissipation functional.  Thus by introducing collective coordinates we can also study the stability of the solutions with $W_0 \neq 0$ in a similar fashion to what we did for the NLSE equation \cite{1751-8121-50-48-485205} by looking at the frequency of the equations linearized around the exact solutions.

%
%
\ack

AK is grateful to Indian National Science Academy (INSA) for awarding him INSA Senior Scientist position at Savitribai Phule Pune University, Pune, India. 
FC and JFD would like to thank the Santa Fe Institute and the Center for Nonlinear Studies at Los Alamos National Laboratory for their hospitality.  FC would also like to thank the Physics Department at Boston University for its hospitality.

%
%
\appendix
%
%
\section{\label{s:Integrals}Useful integrals and definitions}

Definitions:
\begin{subeqnarray}\label{A.e:1}
\fl
   c_1[\gamma]
   &=
   \tint \dd{z} \sech^{\gamma}(z)
   =
   \frac{\sqrt{\pi} \, \GammaF{\gamma/2}}{\GammaF{(\gamma + 1)/2}} \>,
   \label{A.e:1-a} \\
   \fl
   c_2[\gamma]
   &=
   \tint \dd{z} z^2 \, \sech^{\gamma}(z) 
   =
   \frac{2^{\gamma+2}}{\gamma^3} \,
   {}_4F_3\Bigl [\,
      \frac{\gamma}{2},\frac{\gamma}{2},\frac{\gamma}{2},\gamma;
      1+\frac{\gamma}{2},1+\frac{\gamma}{2},1+\frac{\gamma}{2};-1 \Bigr ] \>,
   \label{A.e:1-b} \\
   \fl
   c_3[\gamma]
   &=
   \tint \dd{z} z \, \sech^{\gamma}(z) \, \tanh(z)
   =
   \tint \dd{z} z \, \sech^{\gamma+1}(z) \, \sinh(z) 
   =
   \frac{c_1[\gamma]}{\gamma} \>,
   \label{A.e:1-c} \\
   \fl
   c_4[\gamma]
   &=
   \tint \dd{z} \sech^{\gamma}(z) \, \tanh^2(z)
   =
   c_1[\gamma] - c_1[\gamma+2]
   =
   \frac{c_1[\gamma]}{\gamma+1} \>,
   \label{A.e:1-d}
\end{subeqnarray}
\begin{subeqnarray}\label{A.e:2}
   C_1[\gamma]
   &=
   \frac{c_1^2[\gamma] - c_1^2[\gamma+1]}{c_1[\gamma]\, c_2[\gamma]} \>,
   \label{A.e:2-a} \\
   C_2[\gamma]
   &=
   \frac{c_1[\gamma+1]}{c_1[\gamma]} \>
   \label{A.e:2-b} \\
   C_3[\gamma]
   &=
   \frac{c_1[\gamma+1]}{c_2[\gamma]} \>.
   \label{A.e:2-c}
\end{subeqnarray}
Useful identities:
\begin{subeqnarray}\label{A.e:3}
   c_1[\gamma+2] 
   &= 
   \frac{\gamma}{\gamma+1} \, c_1[\gamma] \>,
   \label{A.e:3-a} \\
   c_2[\gamma+2]
   &=
   \frac{\gamma}{\gamma+1} \, c_2[\gamma] 
   - 
   \frac{2}{\gamma (\gamma+1)} \, c_1[\gamma] \>.
   \label{A.e:3-b}
\end{subeqnarray}
Values:
\begin{eqnarray}\label{A.e:4}
   c_1[1] = \pi \>,
   \qquad
   c_2[2] = 2 \>,
   \qquad
   c_2[1] = \pi^3/4 \>,
   \\
   C_1[1] = 4 (\pi^2 - 4)/\pi^4 = 0.241029 \>.
\end{eqnarray}
We define
\begin{subeqnarray}\label{A.e:5}
   I_1[\gamma,\beta,q]
   &=
   \tint \dd{y} \sech^{\gamma}( \beta y ) \sech(y+q) \>,
   \label{e:I-3a} \\
   I_2[\gamma,\beta,q]
   &=
   \tint \dd{y} \sech^{\gamma}(\beta y) \sech^2(y+q) \>,
   \label{A.e:5-b} \\
   I_3[\gamma,\beta,q]
   &=
   \tint \dd{y} y \sech^{\gamma}( \beta y ) \sech(y+q) \>.
   \label{e:I-3c}
\end{subeqnarray}
We will also need the integrals:
\begin{subeqnarray}\label{I.e:4}
   f_1[\gamma,\beta,q]
   &=
   \tint \dd{y} \sech^{\gamma}(\beta y) \, \sech(y + q) \, \tanh(y + q) \>,
   \label{I.e:4-a} \\
   f_2[\gamma,\beta,q]
   &=
   \tint \dd{y} y \, \sech^{\gamma}(\beta y) \, \sech(y + q) \, \tanh(y + q) \>,
   \label{I.e:4-b} \\
   f_3[\gamma,\beta,q]
   &=
   \tint \dd{y}
   \sech^{\gamma}(\beta y) \, \sech^2(y + q) \, \tanh(y + q) \>,
   \label{I.e:4-c} \\
   f_4[\gamma,\beta,q]
   &=
   \tint \dd{y} \sech^{\gamma}(\beta y) \, \tanh( \beta y) \, \sech(y + q) \>,
   \label{I.e:4-d} \\
   f_5[\gamma,\beta,q]
   &=
   \tint \dd{y} y \,
   \sech^{\gamma}(\beta y) \, \tanh(\beta y) \, \sech(y + q) \>,
   \label{I.e:4-e} \\
   f_6[\gamma,\beta,q]
   &=
   \tint \dd{y} y \,
   \sech^{\gamma}(\beta y) \, \tanh(\beta y) \,
   \sech^2(y + q) \>,
   \label{I.e:4-f} \\
   f_7[\gamma,\beta,q]
   &=
   \tint \dd{y} y^2 \,
   \sech^{\gamma}(\beta y) \, \tanh(\beta y) \, \sech(y + q) \>.
   \label{I.e:4-g} 
\end{subeqnarray}
Partial derivatives of $I_1[\gamma,\beta,q]$ are given by 
\begin{subeqnarray}\label{I.e:5}
\fl
   \pdv{I_1[\gamma,\beta,q]}{q}
   &=
   - \tint \dd{y}
   \sech^{\gamma}(\beta y) \, \sech(y + q) \, \tanh(y + q)
   =
   - f_1[\gamma,\beta,q] \>,
   \label{Ie:5-a} \\
\fl
   \pdv{I_1[\gamma,\beta,q]}{\beta}
   &=
   - \gamma  \tint \dd{y} y \,
   \sech^{\gamma} (\beta y) \, \tanh(\beta y) \, \sech(y + q)
   =
   - \gamma \,  f_{5}[\gamma,\beta,q] \>.
   \label{I.e:5-b} 
\end{subeqnarray}
Partial derivatives of $I_2[\gamma,\beta,q]$ are given by 
\begin{subeqnarray}\label{I.e:6}
\fl
   \pdv{I_2[\gamma,\beta,q]}{q}
   &=
   - 2 \tint \dd{y}
   \sech^{\gamma} (\beta y) \, \sech^2(y + q) \, \tanh(y + q)
   =
   - 2 \, f_3[\gamma,\beta,q] \>,
   \label{I.e:6-a} \\
\fl
   \pdv{I_2[\gamma,\beta,q]}{\beta}
   &=
   - \gamma  \tint \dd{y} y \,
   \sech^{\gamma} (\beta y) \, \tanh(\beta y) \, \sech^2(y + q)
   =
   - \gamma \, f_6[\gamma,\beta,q] \>.
   \label{I.e:6-b} 
\end{subeqnarray}
Partial derivatives of $I_3[\gamma,\beta,q]$ are given by 
\begin{subeqnarray}\label{I.e:7}
\fl
   \pdv{I_3[\gamma,\beta,q]}{q}
   &=
   - \tint \dd{y} y \,
   \sech^{\gamma} (\beta y) \, \sech(y + q) \, \tanh(y + q)
   =
   - f_2[\gamma,\beta,q] \>,
   \label{I.e:7-a} \\
\fl
   \pdv{I_3[\gamma,\beta,q]}{\beta}
   &=
   - \gamma  \tint \dd{y} y^2 \,
   \sech^{\gamma} (\beta y) \, \tanh(\beta y) \, \sech(y + q)
   =
   - \gamma \,  f_7[\gamma,\beta,q] \>.
   \label{I.e:7-b} 
\end{subeqnarray}
Expansion of the integrals about initial values are defined by
\begin{subeqnarray}\label{I.e:8}
   I[\gamma,Q_0+\delta Q]
   &=
   I[\gamma,Q_0] + I_{\mu}[\gamma,Q_0] \, \delta Q^{\mu} + \ldots \>,
   \label{I.e:8-a} \\
   f[\gamma,Q_0+\delta Q]
   &=
   f[\gamma,Q_0] + f_{\mu}[\gamma,Q_0] \, \delta Q^{\mu} + \ldots \>.
   \label{I.e:8-b}
\end{subeqnarray}
We find
\begin{subeqnarray}\label{I.e:9}
\fl
   I_1[\gamma,1+\delta\beta,\delta q]
   &=
   c_1[\gamma+1]
   -
   \gamma \, c_3[\gamma+1] \, \delta \beta + \ldots   
   \label{I.e:9-a} \\
\fl
   &=
   c_1[\gamma+1]
   -
   \frac{\gamma}{\gamma+1} \, c_1[\gamma+1] \, \delta \beta + \ldots \>,
   \notag \\
\fl
   I_2[\gamma,1+\delta\beta,\delta q]
   &=
   c_1[\gamma+2]
   -
   \gamma \, c_3[\gamma+2] \, \delta \beta + \ldots
   \label{I.e:9-b} \\
\fl
   &=
   \frac{\gamma}{\gamma+1} \, c_1[\gamma]
   -
   \frac{\gamma^2}{(\gamma+1)(\gamma+2)} \, c_1[\gamma] \, \delta \beta + \ldots \>,
   \notag \\
\fl
   I_3[\gamma,1+\delta\beta,\delta q]
   &=
   - c_3[\gamma+1] \, \delta q  + \ldots
   =
   - \frac{1}{\gamma+1} \, c_1[\gamma+1] \, \delta q + \ldots \>,
   \label{I.e:9-c}
\end{subeqnarray}
and
\begin{subeqnarray}\label{I.e:10}
\fl
   f_1[\gamma,1+\delta\beta,\delta q]
   &=
   \{\, - c_1[\gamma+1] + 2 c_1[\gamma+3] \,\} \, \delta q + \ldots 
   =
   \frac{\gamma}{\gamma+2} \, c_1[\gamma+1] \, \delta q + \ldots \>,
   \label{I.e:10-a} \\
\fl
   f_2[\gamma,1+\delta\beta,\delta q]
   &=
   c_3[\gamma+1] - \gamma \, c_4[\gamma+1] \, \delta \beta + \ldots 
   \label{I.e:10-b} \\
\fl
   &=
   c_1[\gamma+1]/(\gamma+1)
   -
   \frac{\gamma}{\gamma+2} \,
   \Bigl \{\, 
      \frac{2}{\gamma+1} \, c_1[\gamma+1]
      +
      c_2[\gamma+1] \,
   \Bigr \} \, \delta \beta + \ldots 
   \notag \\
\fl
   f_3[\gamma,1+\delta\beta,\delta q]
   &=
   \{\, - 2 c_1[\gamma+2] + 3 c_1[\gamma+4] \,\} \, \delta q + \ldots
   \label{I.e:10-c} \\
\fl
   &=
   \frac{\gamma^2}{(\gamma+1)(\gamma+3)} \, c_1[\gamma]
   \, \delta q + \ldots \>,
   \notag \\
\fl
   f_4[\gamma,1+\delta\beta,\delta q]
   &=
   \{\, - c_1[\gamma+1] + c_1[\gamma+3] \,\} \, \delta q + \ldots 
   \label{I.e:10-d} \\
\fl
   &=
   -
   \frac{1}{\gamma+2} \, c_1[\gamma+1] \, \delta q + \ldots \>,
   \notag \\
\fl
   f_5[\gamma,1+\delta\beta,\delta q]
   &=
   c_3[\gamma+1] 
   + 
   \{\, - \gamma c_2[\gamma+1] + (\gamma + 1) \, c_2[\gamma+3] \,\} \, \delta \beta + \ldots
   \label{I.e:10-e} \\
\fl
   &=
   \frac{1}{\gamma+1} \, c_1[\gamma+1]
   +
   \frac{1}{\gamma+2} \, \{\, c_2[\gamma+1] - 2 \, c_1[\gamma+1] \,\} \, \delta \beta + \ldots \>,
   \notag \\
\fl
   f_6[\gamma,1+\delta\beta,\delta q]
   &=
   c_3[\gamma+2]
   +
   \{\, - \gamma \, c_2[\gamma+2] +(\gamma + 1) \, c_2[\gamma+4] \,\} \, \delta \beta + \ldots \>,
   \label{I.e:10-f} \\
\fl
   &=
   \frac{\gamma}{(\gamma+1)(\gamma+2)} \, c_1[\gamma]
   +
   \frac{2}{\gamma+3} \,
   \Bigl \{\,
       c_2[\gamma+2] - \frac{\gamma}{\gamma+2} \, c_1[\gamma] \,
   \Bigr \} \, \delta \beta + \ldots \>,
   \notag \\
\fl
   f_7[\gamma,1+\delta\beta,\delta q]
   &=
   \{\, - c_2[\gamma+1] + c_2[\gamma+3] \,\} \, \delta q + \ldots 
   \label{I.e:10-g} \\
\fl
   &=
   - \frac{1}{\gamma+2} \, 
   \Bigl \{\,
      c_2[\gamma+1]
      +
      \frac{2}{\gamma+1} \, c_1[\gamma+1] \,
   \Bigr \} \, \delta q + \ldots \>.
   \notag
\end{subeqnarray}

%
%
\section{\label{s:CCs}Collective coordinates}

The time dependent variational approximation relies on 
 introducing  a finite set of time-dependent real parameters in a  trial wave function that one hopes captures the time evolution of a perturbed solution.  By doing this   one obtains a simplified set of ordinary differential equations for the collective coordinates in place of solving the full partial differential equation for the NLS equation. By judicially choosing the collective coordinates, they can be simply related to the moments of  $x$ and ${\hat p}= -i \partial/\partial x $ averaged over the density $\rho(x,t)$. 
 
That is, we set
\begin{equation}\label{e:VT-1}
   \psi(x,t)
   \mapsto
   \tilde{\psi}[\,x,Q(t)\,] \>,
   \qquad
   Q(t) 
   = 
   \{\, Q^1(t),Q^2(t),...,Q^{2n}(t) \,\} \in \mathbb{R}^{2n} \>.
\end{equation}
The success of the method depends greatly on the choice of the the trial wave function $\tilde{\psi}[\,x,Q(t)\,]$.  The generalized Euler-Lagrange equations lead to Hamilton's equations for the collective coordinates $Q(t)$.
Introducing the notation $\partial_{\mu} \equiv \partial / \partial Q^{\mu}$, the Lagrangian in terms of the collective coordinates is given by
\begin{equation}\label{e:VT-2}
   L(\,Q,\dot{Q}\,)
   =
   \pi_\mu(Q) \, \dot{Q}^\mu - H(\,Q\,) \>,
\end{equation}
where $\pi_\mu(Q)$ is defined by
\begin{equation}\label{e:VT-3}
   \pi_\mu(Q)
   =
   \frac{\rmi}{2} \tint \dd{x}
   \{ \, 
      \tpsi^{\ast}(x,Q)\,[\, \partial_\mu \tpsi(x,Q) \,]
      - 
      [\, \partial_\mu \tpsi^{\ast}(x,Q) \,] \, \tpsi(x,Q)
   \,\} \>,
\end{equation}
and $H(Q)$ is given by
\bea  \label{e:VT-4}
   H(Q)
  && =
   \tint \dd{x} 
   \Bigl \{ \,
       |\partial_x \tpsi(x,Q) |^2   \nonumber \\
     &&  -
      \frac{\rmi \, g}{2(\kappa+1)} 
   \tint \dd{x} \,
   | \tpsi(x,Q) |^{2 \kappa} \,
   [\, \tpsi^{\ast}(x,Q) \, \tpsi_x(x,Q) - \tpsi^{\ast}_x(x,Q) \, \tpsi(x,Q) \, ] \, \nonumber \\
        &&  +
       V_1(x) \, |\tpsi(x,Q)|^2 \,
    \Bigr \} \>. \nonumber \\
\eea
Similarly, in terms of the collective coordinates, the dissipation functional is given by
\begin{equation}\label{e:VT-4.1}
   F[Q,\dot{Q}]
   =
   w_{\mu}(Q) \, \dot{Q}^{\mu} \>,
\end{equation}
where
\begin{equation}\label{e:VT-4.2}
   w_{\mu}(Q)
   =
   \rmi \tint \dd{x} V_2(x) \,
   \{ \, 
      \tpsi^{\ast}(x,Q)\,[\, \partial_\mu \tpsi(x,Q) \,]
      - 
      [\, \partial_\mu \tpsi^{\ast}(x,Q) \,] \, \tpsi(x,Q)
   \,\} \>.
\end{equation}
The generalized Euler-Lagrange equations are
\begin{equation}\label{e:VT-5}
   \pdv{L}{Q^\mu}
   -
   \frac{\partial}{\partial t} \, \Bigl ( \pdv{L}{\dot{Q}^\mu} \Bigr )
   =
   -
   \pdv{F}{\dot{Q}^\mu} \>.
\end{equation}
Setting $v_{\mu}(Q) = \partial_\mu H(Q)$, we find
\begin{equation}\label{e:VT-6}
   f_{\mu\nu}(Q) \, \dot{Q}^\nu
   =
   u_{\mu}(Q)
   =
   v_{\mu}(Q) - w_{\mu}(Q)
\end{equation}
where 
\begin{equation}\label{e:VT-7}
   f_{\mu\nu}(Q)
   =
   \partial_\mu \pi_\nu(Q) - \partial_\nu \pi_\mu(Q)
\end{equation}
is an antisymmetric $2n \times 2n$ symplectic matrix.  
If $\det{f(Q)} \ne 0$, we can define an inverse as the contra-variant matrix with upper indices,
\begin{equation}\label{e:VT-8}
   f^{\mu\nu}(Q) \, f_{\nu\sigma}(Q) = \delta^\mu_\sigma \>,
\end{equation}
in which case the equations of motion \ef{e:VT-6} can be put in the symplectic form:
\begin{equation}\label{e:VT-9}
   \dot{Q}^\mu
   =
   f^{\mu\nu}(Q) \, u_{\nu}(Q) \>.
\end{equation}
Poisson brackets are defined using $f^{\mu\nu}(Q)$.  If $A(Q)$ and $B(Q)$ are functions of $Q$, Poisson brackets are defined by
\begin{equation}\label{e:PB-1}
   \PoissonB {A(Q)}{B(Q)}
   =
   ( \partial_\mu A(Q) ) \, f^{\mu\nu}(Q) \, ( \partial_\nu B(Q) ) \>.
\end{equation}
In particular,
\begin{equation}\label{e:PB-2}
   \PoissonB {Q^\mu}{Q^\nu}
   =
   f^{\mu\nu}(Q) \>.
\end{equation}
It is easy to show that $f_{\mu\nu}(x)$ satisfies Bianchi's identity.  This means that definition \ef{e:PB-1} satisfies Jacobi's identity, as required for symplectic variables.  The rate of energy loss is expressed as
\begin{equation}\label{e:EC-1}
   \dv{H(Q)}{t}
   =
   - v_\mu(Q) \, f^{\mu\nu}(Q) \, w_{\nu}(Q) \>,
   \notag
\end{equation}
since $f^{\mu\nu}(Q)$ is an antisymmetric tensor.

%
%
\section*{References}

%
%
\bibliography{johns.bib}

\providecommand{\newblock}{}
\begin{thebibliography}{10}
\expandafter\ifx\csname url\endcsname\relax
  \def\url#1{{\tt #1}}\fi
\expandafter\ifx\csname urlprefix\endcsname\relax\def\urlprefix{URL }\fi
\providecommand{\eprint}[2][]{\url{#2}}

\bibitem{doi:10.1063/1.523737}
Kaup D~J and Newell A~C 1978 {\em Journal of Mathematical Physics\/} {\bf 19}
  798--801 \urlprefix\url{http://dx.doi.org/10.1063/1.523737}

\bibitem{Ahmed2001343}
Ahmed Z 2001 {\em Phys. Lett. A\/} {\bf 282} 343--348 ISSN 0375-9601
  \urlprefix\url{http://www.sciencedirect.com/science/article/pii/S0375960101002183}

\bibitem{PhysRevE.95.012205}
Chen Y and Yan Z 2017 {\em Phys. Rev. E\/} {\bf 95}(1) 012205
  \urlprefix\url{https://link.aps.org/doi/10.1103/PhysRevE.95.012205}

\bibitem{r:Colin:2006sh}
Colin M and Ohta M 2006 {\em Ann. Inst. H. Poincar{\'e}\/} {\bf AN 23} 753--764

\bibitem{Jenkins2018}
Jenkins R, Liu J, Perry P and Sulem C 2018 {\em Communications in Mathematical
  Physics\/} ISSN 1432-0916
  \urlprefix\url{https://doi.org/10.1007/s00220-018-3138-4}

\bibitem{HayashiOzawa1992}
Hayashi N and Ozawa T 1992 {\em Physica D: Nonlinear Phenomena\/} {\bf 55} 14
  -- 36 ISSN 0167-2789
  \urlprefix\url{http://www.sciencedirect.com/science/article/pii/016727899290185P}

\bibitem{COOPER1993344}
Cooper F, Shepard H, Lucheroni C and Sodano P 1993 {\em Physica D: Nonlinear
  Phenomena\/} {\bf 68} 344 -- 350 ISSN 0167-2789
  \urlprefix\url{http://www.sciencedirect.com/science/article/pii/016727899390129O}

\bibitem{PhysRevLett.71.1661}
Camassa R and Holm D~D 1993 {\em Phys. Rev. Lett.\/} {\bf 71}(11) 1661--1664
  \urlprefix\url{https://link.aps.org/doi/10.1103/PhysRevLett.71.1661}

\bibitem{ROSE1988207}
Rose H~A and Weinstein M~I 1988 {\em Physica D: Nonlinear Phenomena\/} {\bf 30}
  207 -- 218 ISSN 0167-2789
  \urlprefix\url{http://www.sciencedirect.com/science/article/pii/0167278988901078}

\bibitem{COOPER1992184}
Cooper F, Lucheroni C and Shepard H 1992 {\em Physics Letters A\/} {\bf 170}
  184 -- 188 ISSN 0375-9601
  \urlprefix\url{http://www.sciencedirect.com/science/article/pii/037596019291063W}

\bibitem{1751-8121-50-48-485205}
Cooper F, Dawson J~F, Mertens F~G, Ar{\'e}valo E, Quintero N~R, Mihaila B,
  Khare A and Saxena A 2017 {\em Journal of Physics A: Mathematical and
  Theoretical\/} {\bf 50} 485205
  \urlprefix\url{http://stacks.iop.org/1751-8121/50/i=48/a=485205}

\bibitem{doi:10.1063/1.1704233}
Derrick G~H 1964 {\em Journal of Mathematical Physics\/} {\bf 5} 1252--1254
  (\textit{Preprint} \eprint{http://dx.doi.org/10.1063/1.1704233})
  \urlprefix\url{http://dx.doi.org/10.1063/1.1704233}

\bibitem{Vakhitov:1973aa}
Vakhitov N~G and Kolokolov A~A 1973 {\em Radiophys. Quantum Electron\/} {\bf
  16} 783--789

\bibitem{PhysRevE.92.042901}
Kevrekidis P~G, Cuevas-Maraver J, Saxena A, Cooper F and Khare A 2015 {\em
  Phys. Rev. E\/} {\bf 92}(4) 042901
  \urlprefix\url{http://link.aps.org/doi/10.1103/PhysRevE.92.042901}

\bibitem{Liu2013}
Liu X, Simpson G and Sulem C 2013 {\em Journal of Nonlinear Science\/} {\bf 23}
  557--583 ISSN 1432-1467
  \urlprefix\url{https://doi.org/10.1007/s00332-012-9161-2}

\bibitem{PhysRevE.94.032213}
Mertens F~G, Cooper F, Ar\'evalo E, Khare A, Saxena A and Bishop A~R 2016 {\em
  Phys. Rev. E\/} {\bf 94}(3) 032213
  \urlprefix\url{http://link.aps.org/doi/10.1103/PhysRevE.94.032213}

\end{thebibliography}
%
%
\end{document}